%% file: main.tex
\documentclass[journal]{IEEEtran}

\usepackage[noadjust]{cite}
\usepackage{multirow} 
\usepackage{algpseudocode}
\usepackage{algorithm}
\usepackage{rotating}
\usepackage{kantlipsum} 
\usepackage{commath}
\allowdisplaybreaks
\usepackage{mathtools}  
\usepackage{verbatim}
\usepackage{xr-hyper} 
\usepackage{enumitem}

\usepackage{tabularx}

\usepackage{epstopdf}
\usepackage{wrapfig}
\usepackage{latexsym}
\usepackage{amssymb}
\usepackage{amsthm}
\usepackage{amsfonts}
\usepackage{amsmath} 
\usepackage{graphicx}
\usepackage{latexsym}
\usepackage{booktabs}
\usepackage{support-caption} 
\usepackage{subcaption} 
\usepackage{xspace}

\usepackage[normalem]{ulem} 

\usepackage[T1]{fontenc}

\usepackage[dvipsnames]{xcolor}
\usepackage{svg}
\usepackage{caption}
\usepackage{subcaption}

\ifCLASSINFOpdf
\else
\fi
\newcommand{\dq}[1]{``#1''}

\usepackage[dvipsnames]{xcolor}

\newcommand{\commentBy}[3]{\textcolor{#1}{\textbf{#2:} #3}}

\newif\ifcommentson
\commentsontrue

\newcommand{\ste}[1]{\ifcommentson \commentBy{orange}{SS}{#1} \fi}

\newcommand{\cs}[1]{\ifcommentson \commentBy{blue}{CS}{#1} \fi}

\newif\ifextended
\newif\ifshortver

\extendedtrue

\newcommand{\extended}[1]{\ifextended \ifshortver \textcolor{purple}{#1} \else \textcolor{black}{#1} \fi  \fi}
\newcommand{\shortver}[1]{\ifshortver \ifextended \textcolor{blue}{#1} \else \textcolor{black}{#1} \fi \fi}

\newif\ifrevision

\begin{document}

\bstctlcite{IEEEexample:BSTcontrol}


\title{High Performance Delay Monitoring\\for SRv6 Based SD-WANs}
\author{\IEEEauthorblockN{
Carmine Scarpitta\IEEEauthorrefmark{1}\IEEEauthorrefmark{2},
Giulio Sidoretti\IEEEauthorrefmark{1}\IEEEauthorrefmark{2},
Andrea Mayer\IEEEauthorrefmark{1}\IEEEauthorrefmark{2},
Stefano Salsano\IEEEauthorrefmark{1}\IEEEauthorrefmark{2},
\\Ahmed Abdelsalam\IEEEauthorrefmark{4},
Clarence Filsfils\IEEEauthorrefmark{4}\\
}
\IEEEauthorblockA{
\IEEEauthorrefmark{1}University of Rome Tor Vergata,
\IEEEauthorrefmark{2}CNIT,
\IEEEauthorrefmark{4}Cisco Systems
}
\vspace{2ex}
\\ 
\extended{\textbf{Extended version of submitted paper - v02 - December 2022}}
}

\markboth{Submitted Paper}%
{Shell \MakeLowercase{\textit{et al.}}: Bare Demo of IEEEtran.cls for IEEE Journals}
%



\maketitle

\begin{abstract}
Software-Defined Wide Area Networks (SD-WANs) are used to provide services to enterprises with geographically dispersed locations in a flexible and efficient way. We focus on SD-WAN services based on the Segment Routing over IPv6 (SRv6) technology. Performance Monitoring solutions are needed in SD-WANs to detect performance degradation and outages, and optimize network operations.

In this paper, we describe a high performance solution for end-to-end delay monitoring for SRv6 based SD-WAN services. The proposed solution leverages the Simple Two-way Active Measurement Protocol (STAMP) to monitor the delay of an SRv6 path between two nodes called STAMP Session-Sender and Session-Reflector. We describe three implementations of the STAMP Session-Sender and Session-Reflector for a Linux software router and compare their performance. In particular, two implementations are based on user space processing and one is based on eBPF. The results show that the eBPF-based implementation outperforms the user space implementations and has a negligible impact on the forwarding capacity of the Linux software router.

\end{abstract}

\begin{IEEEkeywords}
SD-WAN, Software Defined WAN, Performance Measurement, Segment Routing, SRv6, Delay Monitoring.
\end{IEEEkeywords}

%
\IEEEpeerreviewmaketitle

\input{sec/1-introduction}
\input{sec/2-srv6}
\input{sec/3-sd-wan_srv6}

\input{sec/4-everywan}
\input{sec/5-delay_monitoring_framework}
\input{sec/6-stamp_implementations}
\input{sec/7-delay_monitoring_everywan}
\input{sec/8-monitoring}
\input{sec/9-related_works}
\input{sec/10-conclusions}

\section*{Acknowledgment}
    This work has received funding from the Cisco University Research Program and from the GÉANT Innovation Programme.


%

\extended{
\input{sec/11-appendices.tex}
}


\ifCLASSOPTIONcaptionsoff
  \newpage
\fi



\bibliographystyle{IEEEtran}
\bibliography{references}

\end{document}

%% file: sec/1-introduction.tex
\section{Introduction}
\begin{figure}
    \centering
    \includegraphics[width=\linewidth]{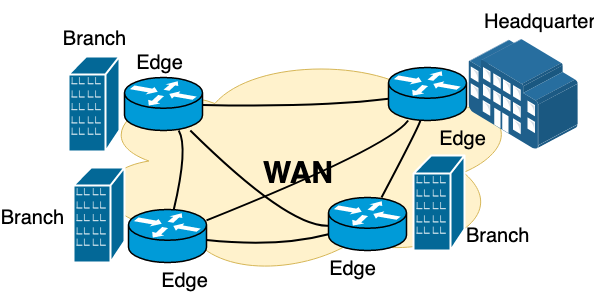}
    \caption{Enterprise WAN reference scenario.}
    \label{fig:sdwan-intro}
\end{figure}

%
%
%
%
\IEEEPARstart{I}{t} is common for enterprises to have multiple data centers and branch offices spread over large geographical areas. The reference scenario is shown in Fig.~\ref{fig:sdwan-intro}. Traditional Wide Area Networks for enterprises were based on static interconnections of remote sites.  
With the advent of cloud computing, many enterprises moved their applications to cloud systems. Traditional WANs started to exhibit limitations because they were not designed for cloud systems. First, traditional WANs do not provide the desired level of flexibility to users. Extending traditional WANs and adding new services require human intervention and are time consuming. Moreover, traditional WANs do not support cloud ecosystems natively. To provide access to cloud applications, traditional WANs typically require backhauling all traffic to a data center. Then, from the data center, the traffic is sent to the cloud.
Software-Defined Wide Area Networking (SD-WAN) is a paradigm that aims at overcoming the limitations of traditional WANs. SD-WAN uses a software-defined approach to control the network and build the interconnections among the different locations. An SD-WAN builds interconnections among users and applications hosted on clouds or remote branches by leveraging any combination of transport services.

Over the years, many SD-WAN solutions have been proposed. Most SD-WAN solutions are commercial, such as Cisco SD-WAN \cite{cisco-sdwan}. The Google B4 WAN \cite{jain2013b4} \cite{hong2018b4} is a proprietary SD-WAN solution that connects Google's data centers across the world. B4 relies on a hybrid SDN approach: the WAN sites are interconnected using traditional routing protocols, an SDN-based Traffic Engineering service runs on top of the network to maximize links utilization and perform load balancing, OpenFlow is used to control and program the switches.
FlexiWAN \cite{flexiwan} was the first open-source solution. It uses VXLAN tunnels to establish the SD-WAN interconnections. In our previous work \cite{scarpitta2021everywan}, we presented an open-source SD-WAN solution called EveryWAN, which is capable of using SRv6 to establish the SD-WAN interconnections. To the best of our knowledge, EveryWAN is the first open-source solution to leverage SRv6 technology to create SD-WAN services. EveryWAN is based on Linux networking and can be deployed on software routers located at the edge of an SD-WAN. 

In fact, software routers can play a role in SD-WAN scenarios, thanks to their flexibility, complementing hardware-based solutions. For example, they can be easily deployed in virtualized environments in cloud and data center scenarios. For this reason, we believe it is fundamental to work on the design and implementation of open-source SD-WAN solutions suitable for software routers. 

An important function to be executed in wide area networks is Performance Monitoring (PM). PM allows network operators to detect failures and outages and assess network performance. Effective network monitoring is essential, and new tools and protocols have been designed accordingly for SDN based networks \cite{tsai2018network}. Important application scenarios in which we can benefit from network monitoring are IoT \cite{bekri2021iot} and security \cite{kebande2020security}.

\extended{To assess performance, there exist several metrics. The \textit{throughput} measures how many packets in the time unit are received. Usually, it is measured in bit/sec. \textit{Packet loss} is a measure of how many packets fail to reach the destination due to network congestion or transmission errors. \textit{Delay} (or \textit{latency}) is a measure of the time required by a packet to reach the destination, and it is usually expressed in milliseconds. There are two types of delay: i) the \textit{one way} delay is the delay measured from the source to the destination; ii) the \textit{two way} delay is the delay measured from the source to the destination plus the delay measured from the destination to the source. Finally, the \textit{jitter} measures the variation of the delay.}

\extended{Moreover, the measurement methods can be classified in \textit{passive methods}, \textit{active methods} or \textit{hybrid}. \textit{Passive} methods typically use a sniffer to analyze the traffic flows in real-time and extract the relevant statistics. Passive methods do not alter the traffic flows, but are usually less effective than active methods. On the contrary, \textit{active} methods typically inject probe packets in the network that carry useful information for monitoring purposes. Finally, \textit{hybrid} methods try to combine the advantages of active and passive methods.}

In this paper, we focus on the delay monitoring of SRv6 networks. We consider a number of research and technological questions:
\begin{itemize}
  \item Is it possible to design an effective solution for delay monitoring of SRv6 networks based on current IETF standards and work-in-progress Internet drafts?
  \item Can we implement the solution in a working open-source prototype based on Linux software routers?
  \item What is the impact of the delay monitoring solutions on the forwarding capacity of software routers? Can we implement delay monitoring with negligible impact on forwarding performance?
\end{itemize}

The main novel contributions are as follows:

\begin{itemize}
  \item Realization of a High Performance End-to-End Delay Monitoring solution for SRv6 networks compliant with available standards and Internet drafts;
  \item Design and implementation of a gRPC Southbound interface to control the SRv6 nodes;
  \item Implementation of two user space solutions and of a kernel solution based on eBPF;
  \item Evaluation of the performance degradation introduced by the Delay Monitoring solution and comparison between the two user space and the eBPF-based implementations.
\end{itemize}

This paper is organized as follows. In Section \ref{sec:srv6}, we present an introduction to the SRv6 technology and its main use cases. Section \ref{sec:sd-wan_srv6} presents how the SRv6 technology can be used to realize SD-WAN services. In Section \ref{sec:everywan}, we introduce EveryWAN, the SD-WAN prototype that we have extended. Section \ref{sec:delay-monitoring-framework} presents our Delay Monitoring solution. The implementations are discussed in Section \ref{sec:stamp_implementations}. In Section \ref{sec:delay_monitoring_everywan}, we show how we integrated our performance measurement solution in EveryWAN to measure the delay of VPN services. In Section \ref{sec:results}, we present a performance evaluation and comparison of the implementations. In section \ref{sec:related-works} we present the related works. Finally, Section \ref{sec:conclusion} concludes the paper.

%% file: sec/2-srv6.tex
\section{SRv6 Technology}
\label{sec:srv6}

Segment Routing (SR) is a routing technology based on the \textit{loose source routing} paradigm (\cite{filsfils2015srv6arch}, \cite{rfc8402srv6arch}). It allows a source node to steer a packet through a list of instructions called \textit{segments}. A segment can represent a topological instruction (e.g., forward the packet via a specific nexthop) or a function to be applied to the packet (e.g., execute an operation on the packet). A segment is identified by an identifier known as \textit{Segment ID (SID)}. The list of SIDs of a packet, called \textit{Segment List} or \textit{SID List}, is carried in the packet header. SR can be implemented using either MPLS or IPv6 as data plane technology. In MPLS Segment Routing (SR-MPLS) \cite{rfc8663mpls}, the SIDs are encoded as MPLS labels. The Segment List is encoded as a stack of labels. In Segment Routing over IPv6 (SRv6), the SIDs are encoded as IPv6 addresses. The Segment List is carried in an IPv6 Extension Header called Segment Routing Header (SRH) \cite{rfc8754srh}. A set of standardized SRv6 functions is presented in \cite{rfc8986netprog}. In this paper, we focus on SRv6. 

\begin{figure}
    \centering
    \includegraphics[width=\linewidth]{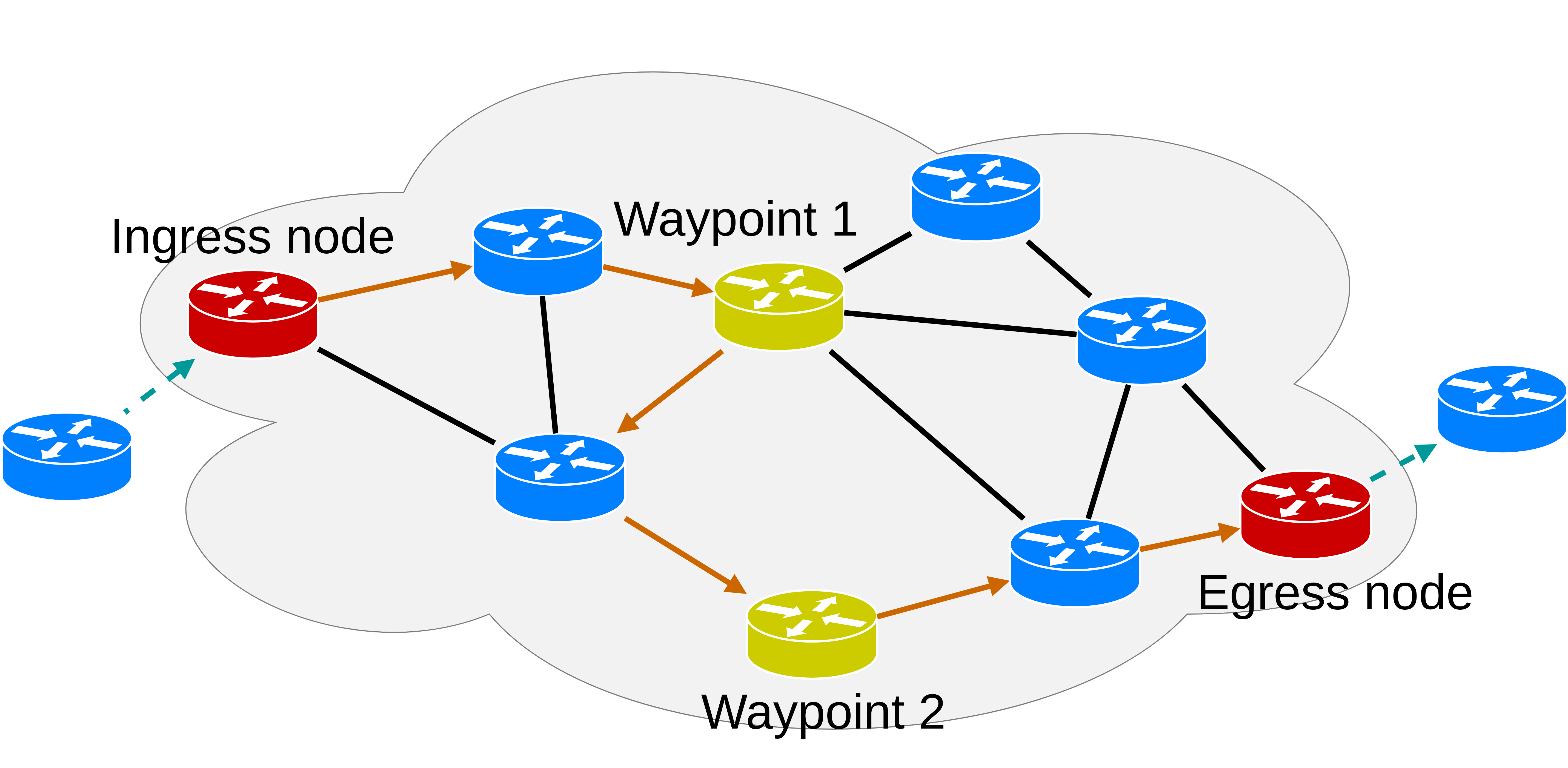}
    \caption{SRv6 example network scenario.}
    \label{fig:srv6-data-plane}
\end{figure}

Fig.~\ref{fig:srv6-data-plane} shows an example of an SRv6 network scenario. The gray cloud represents an SRv6 domain. An ingress node processes the packets entering the SRv6 domain and encapsulates each received packet in an outer IPv6 header with an SRH. In the example, the SRH carries a SID List containing three SIDs. The first two SIDs represent the two waypoints that the packets should traverse before reaching the destination. The ingress node forwards the encapsulated packets towards the first waypoint. The path to reach the waypoint is decided by the traditional routing protocol (e.g., IS-IS or OSPF). The first waypoint forwards the packet toward the second waypoint, which in turn forwards the packet toward the egress node, identified by the third SID in the segment list. The third SID is also used in the egress node to determine the operation to be performed. In this case, the egress node performs a decapsulation operation (i.e., removes the outer IPv6 header which contains the SRH) and forwards the packets to the destination. It is also possible to use two different SIDs instead of the third single one: a SID to reach the egress node and another SID to identify the operation to be performed, but this is less efficient as four SIDs instead of three would be carried in the Segment List.

The SRv6 technology has been proposed in the recent past and has raised great interest in academia and industry. Since then, its development has progressed very rapidly. Today, SRv6 is supported in many hardware deployments \cite{matsushima2022srv6deployment} and software routers such as the Linux kernel and the Vector Packet Processor (VPP) \cite{vpp}. The Linux kernel has supported SRv6 packet generation and forwarding capabilities since version 4.10 (released in February 2017). Later, it has been extended to support many of the SRv6 behaviors described in \cite{rfc8986netprog}.

SRv6 enables many use cases such as overlay VPNs \cite{ietf-bess-srv6-services-15}, Traffic Engineering \cite{rfc7855}, Fast Rerouting \cite{rfc7855}, and Service Function Chaining (SFC) \cite{li-spring-sr-sfc-control-plane-framework-06}. An overview of SRv6 implementation and deployment status is available at \cite{ventre2021survey} and \cite{matsushima2022srv6deployment}. The ROSE project \cite{roseproject} aims to build a Linux-based Open Ecosystem for SRv6. It tackles multiple aspects of the SRv6 technology, including the Data Plane, Control Plane, SRv6 host networking stack, integration with applications, and integration with Cloud/Data Center Infrastructures. \shortver{ROSE comprises several sub-projects which are the foundation of the work presented here.}\extended{ROSE comprises several sub-projects, including SRPerf \cite{abdelsalam2021srperf} (a performance evaluation framework for SRv6 implementations), SRv6-PM \cite{loreti2021srv6pm} (a loss monitoring solution for SRv6 networks), and HIKe \cite{mayer2021hike} (a solution that combines the advantages of Linux kernel networking and custom-designed eBPF programs to speed up the performance of SRv6 software routers).
In \cite{tulumello2020microsids}, the authors proposed a solution to efficiently represent SIDs, called \textit{Micro SID}. This solution reduces the length of the SID List and facilitates using SRv6 on devices with limited processing capabilities.}

%% file: sec/3-sd-wan_srv6.tex
\section{SD-WAN Services Based on SRv6}
\label{sec:sd-wan_srv6}

In the UCSS project\footnote{part of the GÉANT Innovation Programme \cite{geant-innovation-programme}} (\textit{User Controlled SD-WAN Services with Performance Monitoring over GÉANT)} \cite{ucss-report} we designed, implemented and deployed SD-WAN services based on the SRv6 technology.

\begin{figure}
    \centering
    \includegraphics[width=\linewidth]{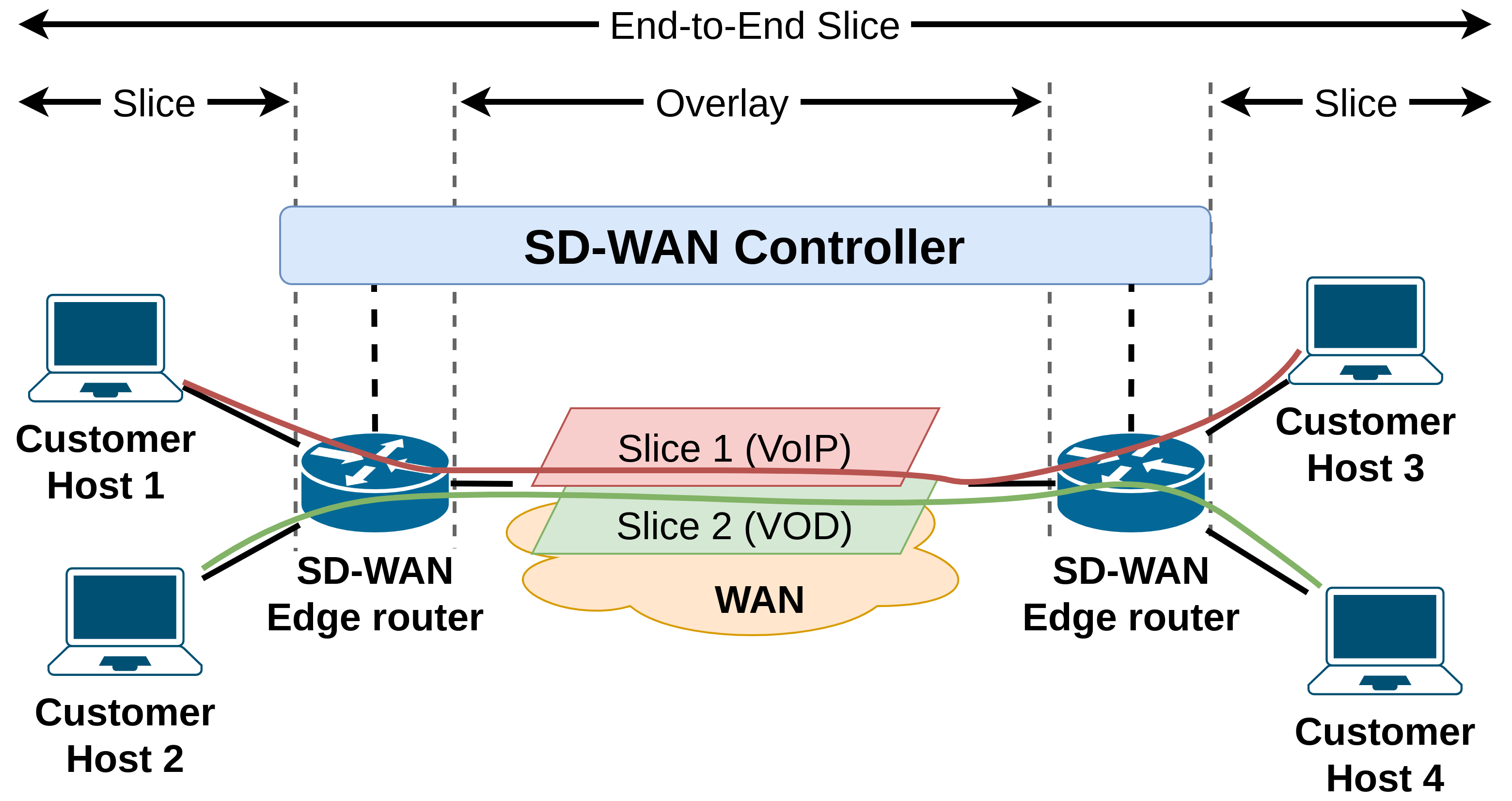}
    \caption{SD-WAN Network Slicing scenario.}
    \label{fig:network-slicing}
\end{figure}

An SD-WAN can offer different services. We focus on the \textit{Network Slicing} service. The reference scenario is shown in Fig.~\ref{fig:network-slicing}. Network Slicing allows customers to create different logical instances of virtual networks over the same WAN connection. It allows multiple applications to run in isolation over the same WAN. Among the different types of slicing, we focus on \textit{Routed End-to-End Slices}. A Routed End-to-End Slice is an implementation of a Layer 3 VPN (L3VPN), in which the devices attached to the SD-WAN Edges belong to different broadcast domains. The SD-WAN Edge routers act as gateways to route the traffic between these broadcast domains.

\begin{figure*}
    \centering
    \includegraphics[width=0.8\linewidth]{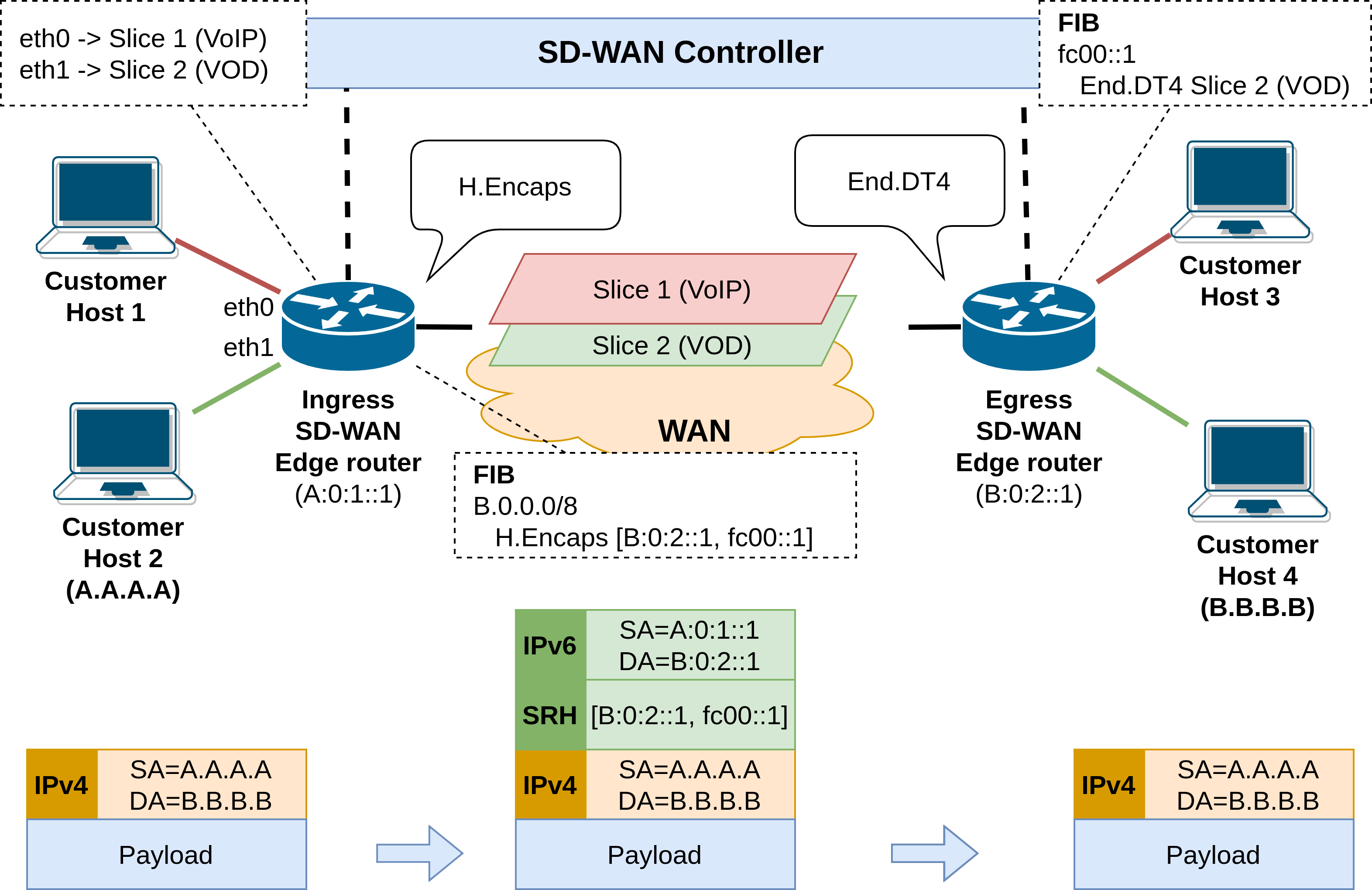}
    \caption{SD-WAN Network Slicing scenario (SRv6).}
    \label{fig:srv6-network-slicing}
\end{figure*}

In our terminology, a \textit{Slice} (or \textit{Local Slice}) is a portion of the customer network where users or applications are located. Each Local Slice is terminated in an SD-WAN Edge router. The SD-WAN Edge router forwards the traffic of the connected Local Slice to an egress SD-WAN Edge router. The interconnections between two different SD-WAN Edge routers are realized by using a set of \textit{Tunnels} (also called \textit{Overlays}). The Overlay, together with the Local Slices, forms the so-called \textit{End-to-End Slice} (\textit{E2E Slice}). Several technologies can be used to realize an Overlay. We focus on SRv6-based Overlays. Fig.~\ref{fig:srv6-network-slicing} shows the reference scenario for the Network Slicing service based on SRv6 technology. An \emph{ingress} SD-WAN Edge router receives IP packets from a customer source host. It classifies and associates each incoming packet with a specific End-to-End Network Slice according to various criteria, such as the incoming interface, the source IP address, or the protocol. After the classification, the ingress SD-WAN Edge router performs a lookup in its Forwarding Information Base (FIB) to discover the SD-WAN \emph{egress} Edge router attached to the destination host. Then, the ingress SD-WAN Edge router applies the \textit{H.Encaps} behavior described in \cite{rfc8986netprog} to the packet. This behavior steers the packet into an SRv6 Policy. Steering is realized by encapsulating the IP packet into an outer IPv6 header containing an SRH. The SRH carries two SIDs. The first SID represents an instruction to deliver the packet to the egress SD-WAN Edge router. The second SID is an \textit{End.DT6} instruction. End.DT6 forces the egress router to strip the outer IPv6+SRH header and deliver the original packet to the correct Slice.


In the SD-WAN solutions, the \textit{SD-WAN Edge routers} are deployed in all the locations where the SD-WAN interconnections need to be established. An \textit{SD-WAN controller} manages and programs the SD-WAN Edge routers. Depending on the location and the characteristics of the SD-WAN Edge routers, three scenarios are possible:
\begin{enumerate}
    \item the SD-WAN Edge routers are located within the provider network and are under the network operator's control;
    \item the SD-WAN Edge routers are outside the provider network, and they have no control over the transport services;
    \item the SD-WAN Edge routers are outside the provider network but can interact with the provider network to deploy the SD-WAN services.
\end{enumerate}

We focused on scenario 2. The SRv6-based SD-WAN services were deployed in scenarios where the SD-WAN Edge routers have no interaction with the provider network. We deployed several SD-WAN Edge routers as Virtual Machines (VMs) across Europe. These SD-WAN Edge routers were located in different kinds of networks, like university campus networks,  NRENs (National Research and Education Networks), and commercial provider networks. We analyzed and classified the IPv6/SRv6 connectivity between these VMs and introduced the concept of \textit{SRv6 Transparency}. SRv6 Transparency is the ability of an IPv6 network to carry SRv6 traffic. Several factors can reduce the SRv6 Transparency of a network, such as firewalls that block IPv6 packets carrying an SRH. We found different SRv6 Transparency levels in the networks that we considered. We have shown that it is possible to configure the SRv6-based SD-WAN services, taking into account the SRv6 Transparency level of the network providing IPv6 connectivity and we have practically deployed SD-WAN services across operational networks over the Internet. 
An in-depth discussion of the SRv6 Transparency problem and the configuration of SRv6-based SD-WAN services can be found in the UCSS report \cite{ucss-report}.

%% file: sec/4-everywan.tex
\section{The EveryWAN Architecture}
\label{sec:everywan}

\begin{figure}
    \centering
    \includegraphics[width=\linewidth]{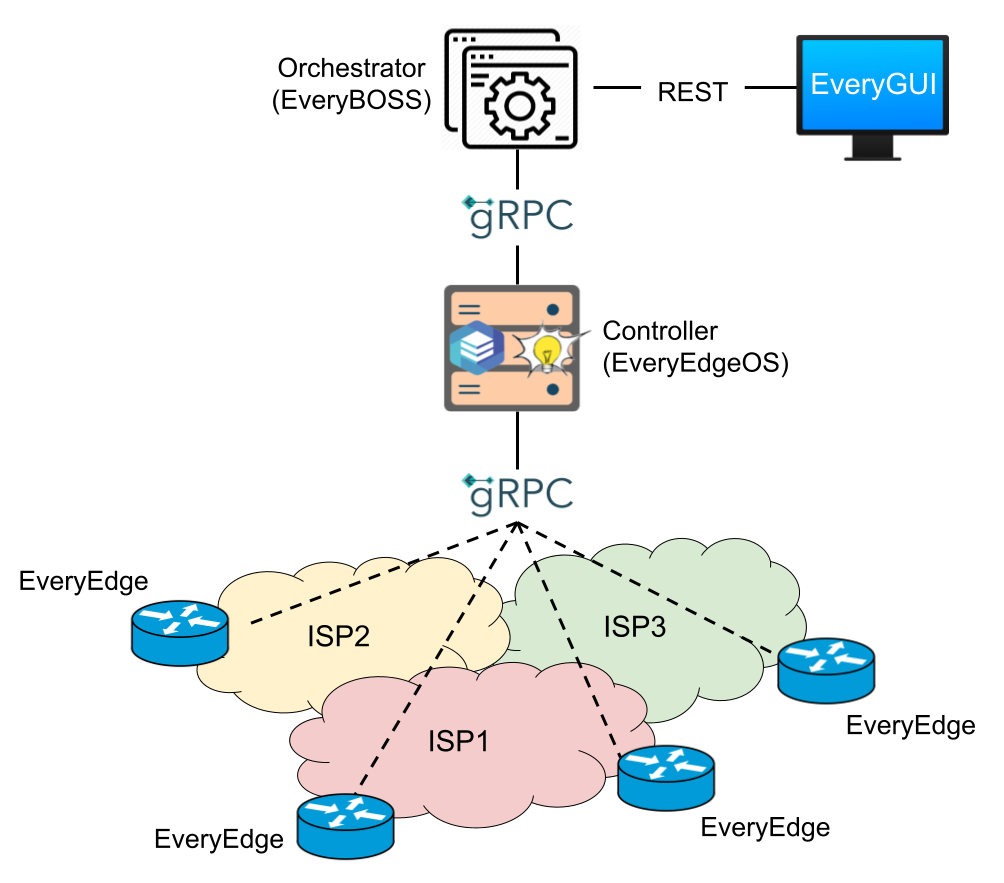}
    \caption{EveryWAN Architecture.}
    \label{fig:everywan-arch}
\end{figure}



\textit{EveryWAN} \cite{scarpitta2021everywan} is an open-source SD-WAN prototype based on Linux networking. Fig.~\ref{fig:everywan-arch} shows the EveryWAN architecture. At the lowest level, we have the SD-WAN Edge routers called \textit{EveryEdge routers}. The EveryEdge routers take care of the interconnections among all the sites. 
EveryEdge routers can be deployed as \textit{Virtual Network Functions (VNFs)} over a Linux OS in the sites to be interconnected. An SD-WAN Controller, called \textit{EveryEdgeOS}, manages all the EveryEdge routers through an API based on the gRPC protocol. It deals with many configuration and management aspects of the EveryEdge routers, ranging from their initial registration, authentication, and configuration to the activation of the policies that implement the SD-WAN services. On top of the controller, there is an SD-WAN Orchestrator named \textit{EveryBOSS}, which automates the deployment of the EveryEdge routers and SD-WAN services. The orchestrator also offers a GUI that allows the customers to configure the EveryEdge routers and manage the SD-WAN services. The EveryEdgeOS and the EveryBOSS orchestrator can run either in a self-managed private cloud or in a public cloud.

The EveryEdge router comprises several open-source components installed on a general-purpose Linux distribution (e.g., Ubuntu Server). It uses Linux networking capabilities to forward the traffic. A component called \textit{EveryEdgeManager} offers a \textit{Southbound API} that allows the EveryEdgeOS controller to program and configure the router. Through the Southboud API, the controller can send commands to the EveryEdge router (e.g., install a specific route or set the IP address of a network interface). The EveryEdgeManager translates the received commands into lower-level actions. Then, it sends these actions to the Linux kernel using the open-source project \textit{pyroute2} \cite{pyroute2}, a pure Python \textit{Netlink} library. A detailed description of the EveryEdge router architecture can be found in \cite{scarpitta2021everywan}.

The main service offered by EveryWAN is Network Slicing (described in Section \ref{sec:sd-wan_srv6}), which allows customers to create End-to-End Slices among the remote sites. \extended{To create End-to-End Slices, the EveryEdge routers must be configured properly. The configuration depends on the SRv6 Transparency of the network.}
The EveryEdge router receives ingress IP packets over the customer-facing interfaces, i.e., the Local interfaces (LAN). It classifies and associates each packet with a particular End-to-End Network Slice. To perform the classification, the EveryEdge leverages the \textit{Virtual Routing and Forwarding} (VRF) technology offered by the Linux kernel. VRFs provide the ability to create isolated virtual routing and forwarding domains. Each VRF serves a particular slice. Each customer-facing interface in the EveryEdge router is mapped to a slice and enslaved to the VRF that serves that slice. Based on the destination IP address, the EveryEdge router forwards the packets associated with a slice to the remote EveryEdge routers over the WAN interfaces.

A transport technology ensures that the network delivers the packets to the remote EveryEdge router. EveryWAN supports two transport technologies: VXLAN \cite{rfc7348} and SRv6. In this work, we only consider SRv6.
\shortver{
To transmit the packets using SRv6, the EveryEdge routers use the \textit{H.Encaps} and the \textit{End.DT4/End.DT6} behaviors as depicted in Fig. \ref{fig:network-slicing} and discussed in the previous section.}
\extended{
To transmit the packets to a remote router, the EveryEdge router (which is an ingress EveryEdge router from the SRv6-domain point of view) applies the \textit{H.Encaps} behavior described in \cite{rfc8986netprog} to the incoming packets. This behavior steers the packets into an SRv6 Policy. Steering is realized by encapsulating the IP packets into an outer IPv6 packet containing an SRH. The SRH carries two SIDs. The first SID represents an instruction to deliver the packet to the remote EveryEdge router. The second SID is an instruction End.DT4/End.DT6 that strips the outer IPv6+SRH header and delivers the original packet to the VRF that serves the slice.
The encapsulated packets traverse the network and reach the remote EveryEdge router (which acts as an egress EveryEdge router). The egress EveryEdge router receives the SRv6 packets over the WAN interfaces. It applies the \textit{End.DT4}/\textit{End.DT6} behavior to decapsulate the SRv6 packet and forward the original IP packets over the LAN interfaces to the destination.
}

A detailed description of the EveryWAN architecture can be found in the white paper \cite{everywan-report}.

%% file: sec/5-delay_monitoring_framework.tex
\section{STAMP Delay Monitoring for SRv6}
\label{sec:delay-monitoring-framework}




In this section, we present the proposed End-to-End Delay Monitoring solution for SRv6 networks based on the Simple Two-Way Active Measurement Protocol (STAMP) \cite{rfc8762stamp}. STAMP enables the measurement of several performance metrics, including \textit{packet loss}, \textit{delay}, and \textit{jitter}. It supports both one-way and round-trip measurements in IP networks. RFC 8762 \cite{rfc8762stamp} defines the base functionalities of STAMP and describes the format of the packets that collect and carry the measurement data. RFC 8972 \cite{rfc8972stampext} introduces the \textit{STAMP Session IDentifier} (\textit{SSID}) and defines optional STAMP extensions that enhance the STAMP base functions. The drafts \cite{ietf-ippm-stamp-srpm-03} and \cite{ietf-spring-stamp-srpm-03} present general guidelines for measuring various performance metrics in SR networks using STAMP. In the following subsections, we present a solution based on STAMP to measure the end-to-end delay of SRv6 paths.

\begin{figure}[t!]
  \centering
  \includegraphics[width=0.9\linewidth]{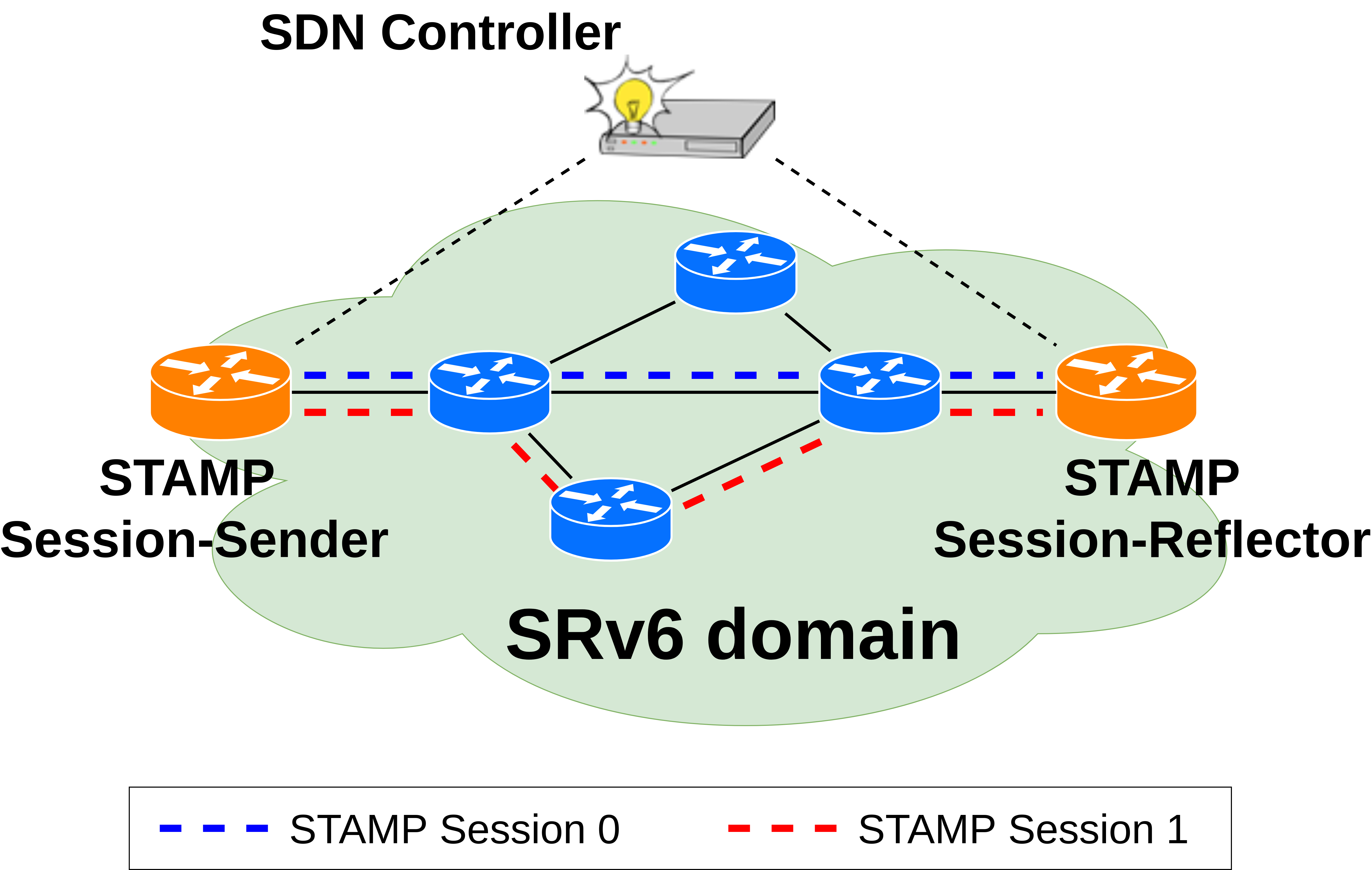}
  \caption{STAMP reference scenario.}
  \label{fig:stamp-reference-scenario}
\end{figure}

Fig.~\ref{fig:stamp-reference-scenario} shows our STAMP reference scenario. We use a \textit{STAMP Session} to measure the end-to-end delay on an SRv6 path between two nodes called \textit{STAMP Session-Sender} and \textit{Session-Reflector}. For delay measurements to be meaningful, the Session-Sender and Session-Reflector clocks must be synchronized\footnote{Clock synchronization mechanisms are out of scope for this paper, we assume that the clocks are synchronized.}. RFC 8762 does not envisage any particular approach to configure and manage the STAMP Session-Sender, Session-Reflector, and the STAMP Session, which can be achieved in different ways, such as using a Command Line Interface (CLI) or an SDN controller. The proposed solution leverages an SDN controller to manage the STAMP Session and configure the STAMP Session-Sender and Session-Reflector. The public documentation of our delay monitoring solution with links to code repositories is available in~\cite{srv6-delay-mon}.

\subsection{Data Plane Protocol}
\label{sec:delay-monitoring-framework-dataplane}

A STAMP session measures the end-to-end delay on a given SRv6 path between two nodes, the STAMP Session-Sender and Session-Reflector. A STAMP session consists of a bidirectional packet exchange between the STAMP Session-Sender and the Session-Reflector. Each STAMP session is identified by a unique 16-bit nonzero unsigned integer called \textit{STAMP Session IDentifier} (\textit{SSID}).

\begin{figure}
  \centering
  \begin{subfigure}[htbp]{0.9\linewidth}
    \centering
    \includegraphics[width=\linewidth]{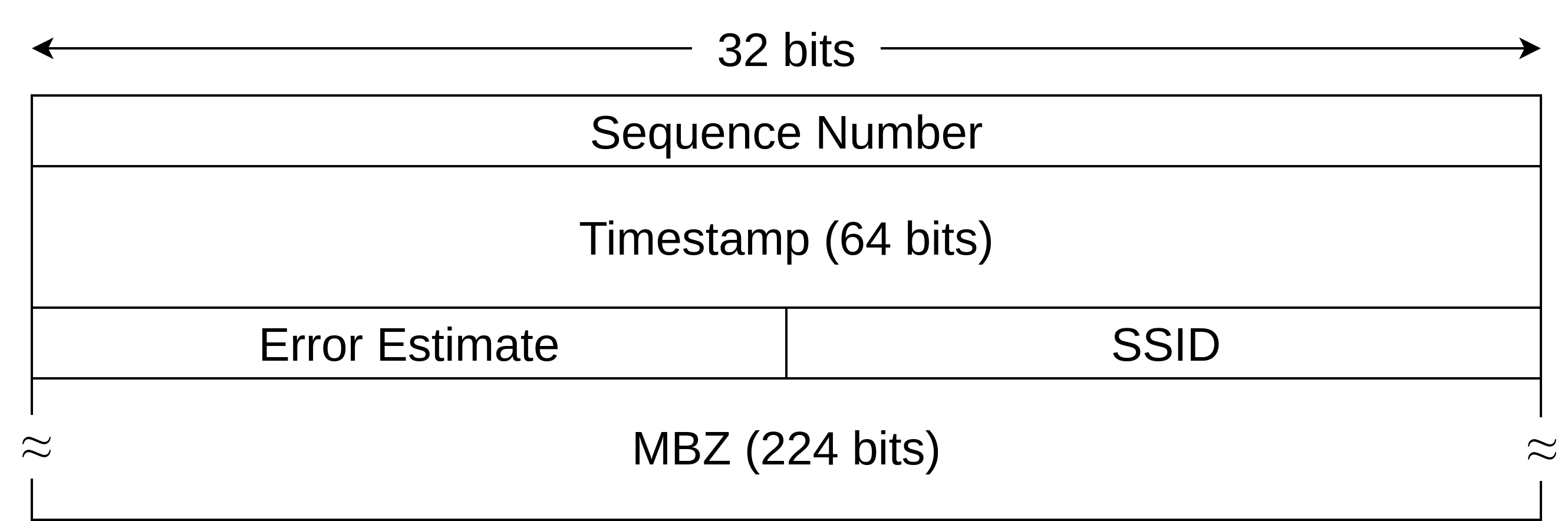}
    \caption{STAMP Session-Sender Test Packet.}
    \label{fig:stamp-sender-pkt}
  \end{subfigure}
  \par\bigskip
  \begin{subfigure}[htbp]{0.9\linewidth}
    \centering
    \includegraphics[width=\linewidth]{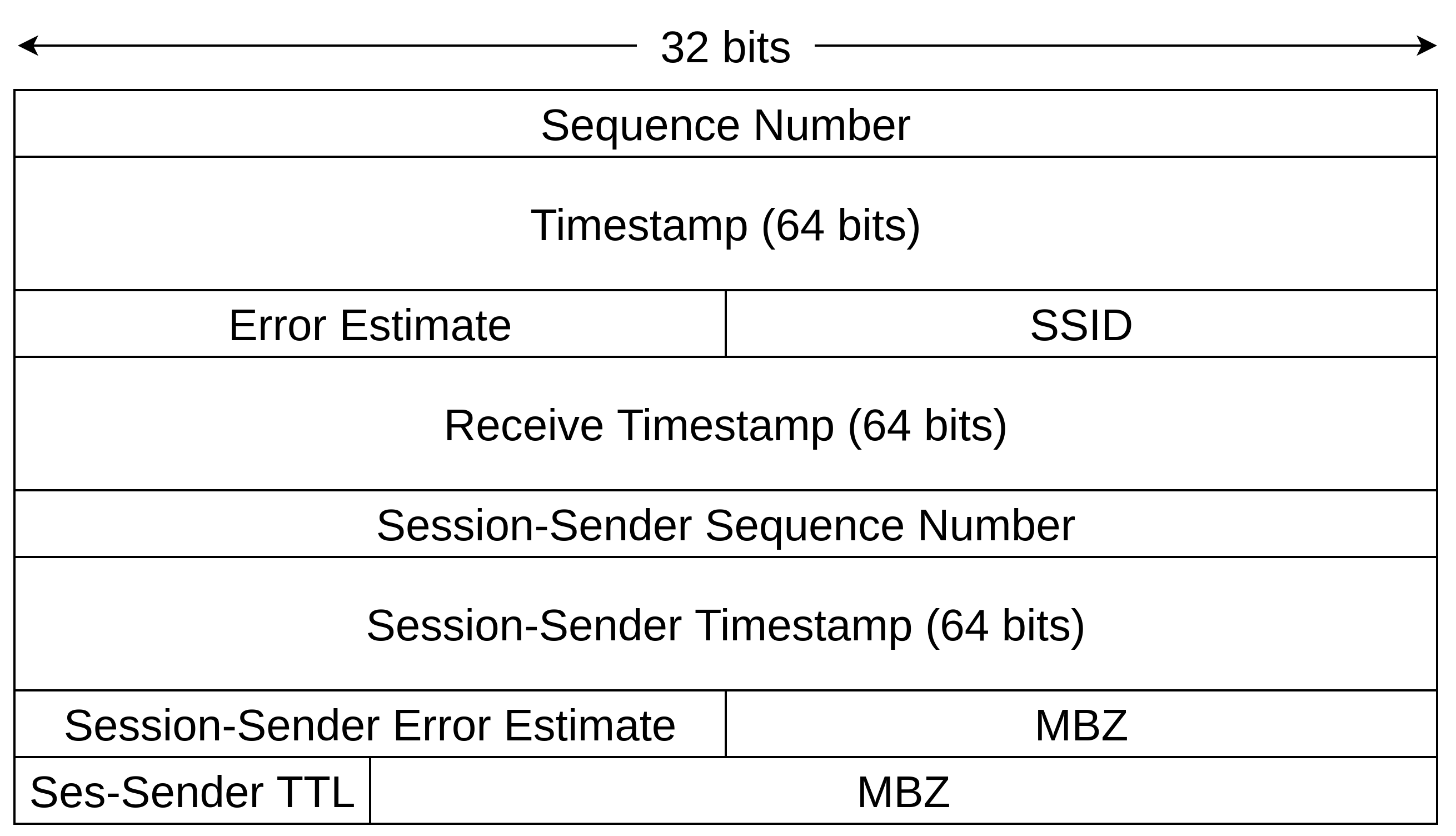}
    \caption{STAMP Session-Reflector Test Packet.}
    \label{fig:stamp-reflector-pkt}
  \end{subfigure}
  \caption{STAMP Test Packets defined in \cite{rfc8972stampext}.}
\end{figure}

\begin{figure}[t!]
  \centering
  \includegraphics[width=0.9\linewidth]{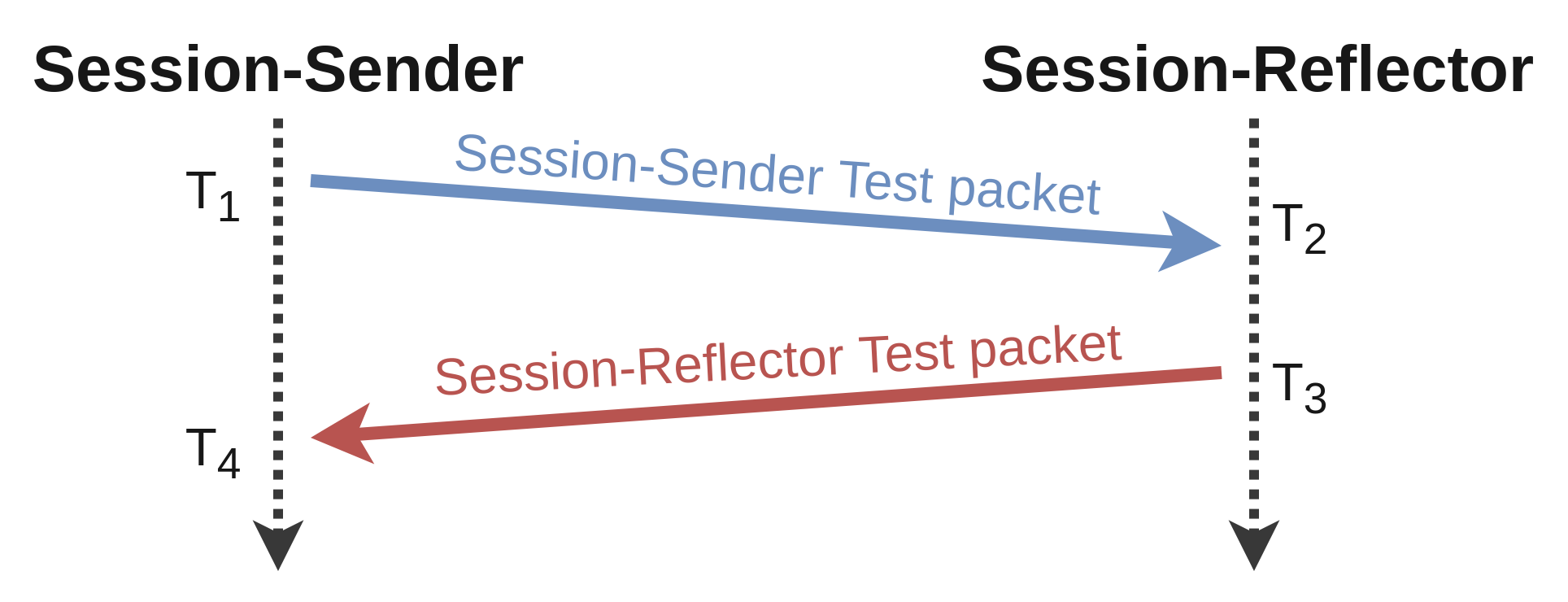}
  \caption{STAMP time diagram.}
  \label{fig:stamp-time-diagram}
\end{figure}

The STAMP Session-Sender transmits a \textit{STAMP Session-Sender test packet} to the STAMP Session-Reflector. The test packet is an IPv6/UDP packet sent to the STAMP UDP port of the Session-Reflector. By default, the STAMP Session-Reflector uses the UDP port 862. The SDN controller can set a different port during the configuration of the STAMP Session-Reflector. The STAMP Session-Sender test packet is transmitted on the same path as the data traffic flow under measurement to measure the delay experienced by the data traffic flow. To enforce the path, the STAMP Session-Sender adds an SRH to the IPv6 header. The SRH contains the SID List that encodes the path under measurement from the STAMP Session-Sender to the Session-Reflector. The test packet carries the payload shown in Fig.~\ref{fig:stamp-sender-pkt}. The \textit{Sequence Number} field contains a 32-bit unsigned integer. It starts at zero and is incremented by one with each sent packet. The \textit{Timestamp} field carries the time when the Session-Sender sent the test packet. In the rest of this section, we refer to this timestamp as $T_1$ (see Fig.~\ref{fig:stamp-time-diagram}). RFC 8762 specifies two different timestamp formats: \textit{Network Time Protocol} (NTP) \cite{rfc5905} and the IEEE 1588v2 \textit{Precision Time Protocol} (PTP) \cite{ieee1588}, both using 64 bits. By default, the STAMP Session-Sender uses NTP as timestamp format, as specified in RFC 8762. The SDN controller can select a different timestamp format during the STAMP Session-Sender or STAMP Session configuration. 

\extended{The \textit{Error Estimate} field comprises four sub-fields: i) the \textit{S} field; ii) the \textit{Z} field; iii) the \textit{Scale} field; iv) the \textit{Multiplier} field. S is a one-bit field that indicates whether the STAMP Session-Sender has a clock that is synchronized to UTC using an external source like GPS hardware or not. The one-bit Z field indicates the timestamp format used for the test packet (i.e., NTP or PTPv2). Scale and Multiplier provide an error estimate.}
The \textit{SSID} (Session Sender ID) field contains the SSID of the STAMP Session to which the test packet belongs. It associates the STAMP Session-Sender test packet with the corresponding STAMP Session. The remaining 28 bytes (224 bits) are set to zero (\textit{Must-Be-Zero} or \textit{MBZ} field). The content of STAMP Session-Reflector test packet is larger than the content of a STAMP Session-Sender test packet. The MBZ field makes the size of the Session-Sender test packet equal to the size of the Session-Reflector test packet.

Following the SRv6 path under measurement, the test packet is delivered to the Session-Reflector. The Session-Reflector receives the STAMP Session-Sender test packet and verifies it. If the packet is valid and the SSID corresponds to an active STAMP Session, the Session-Reflector creates and sends a STAMP Session-Reflector test packet to the STAMP UDP port of the Session-Sender. The STAMP Session-Reflector test packet carries the payload depicted in Fig.~\ref{fig:stamp-reflector-pkt}. Bytes 24-33 contain an exact copy of the STAMP Session-Sender test packet. The \textit{Sequence Number} field contains a 32-bit unsigned integer. The STAMP Session-Reflector can work in two modes: i) \textit{stateless} mode; ii) \textit{stateful} mode. In the stateless mode, the STAMP Session-Reflector reuses the same Sequence Number value contained in the STAMP Session-Sender test packet. In the stateful mode, the STAMP Session-Reflector maintains a counter for the transmitted packets. \extended{This counter starts at zero and is incremented for each transmitted packet in the context of the STAMP Session and used as Sequence Number field in the packets transmitted by the Session-Reflector.}The \textit{Receive Timestamp} field contains the time when the Session-Reflector received the Session-Sender test packet, denoted as $T_2$  (see Fig.~\ref{fig:stamp-time-diagram}). The \textit{Timestamp} field contains the time when the Session-Reflector starts transmitting the Session-Reflector test packet, denoted as $T_3$. \extended{Similarly to the Timestamp field of the STAMP Session-Sender test packet, the Timestamp and Receive Timestamp can be encoded using either NTP or PTPv2.}\extended{The \textit{Error Estimate} field indicates the synchronization bit, timestamp format and error estimation of the Timestamp and Receive Timestamp and has the same structure as the Error Estimate field in the Session-Sender test packet.}The \textit{SSID} 16-bit field contains the STAMP Session IDentifier and allows the STAMP Session-Sender to associate the received STAMP Session-Reflector packets with the correct STAMP Session. The \textit{Session-Sender TTL} is a copy of the Hop Limit field of the IPv6 header contained in the received STAMP Session-Sender test packet. The \textit{MBZ} fields are used to achieve an alignment on a four-byte boundary. The Session-Reflector test packet is transmitted on the same path as the data traffic flow under measurement to measure the delay experienced by the data traffic flow. This can be the same path as the Session-Sender test packet or a different path. The draft \cite{ietf-ippm-stamp-srpm-03} defines a TLV called \textit{Return Path TLV} that allows the Session-Sender to request the Session-Reflector to transmit the Session-Reflector test packet on a specific path. However, we do not use the Return Path TLV in our solution. We leverage the SDN controller to set up the return path as part of the STAMP Session configuration. Before sending the STAMP Session-Reflector test packet, the Session-Reflector adds an SRH to the IPv6 header to enforce the return path. The SRH contains a SID List that encodes the path under measurement from the STAMP Session-Reflector to the Session-Sender.

Following the path specified in the SID List, the STAMP Session-Reflector test packet is delivered to the Session-Sender. The Session-Sender verifies the packet and validates the SSID. If the SSID corresponds to an active STAMP Session, it generates a new timestamp $T_4$, which is the time when the Session-Sender received the Session-Reflector test packet. The Session-Sender collects the three timestamps from the session reflector test packet and adds $T_4$ creating a measurement record ($T_1$, $T_2$, $T_3$, $T_4$) that is stored locally. The generated records need to be sent to the SDN controller for post-processing, as it will be discussed later. Considering its role in the processing of the STAMP test packets coming back from the Session-Reflector, we can refer to the Session-Sender as the final \emph{Collector} of the STAMP test packets.


\subsection{Configuration and Management}

\begin{figure}[t!]
  \centering
  \includegraphics[width=0.9\linewidth]{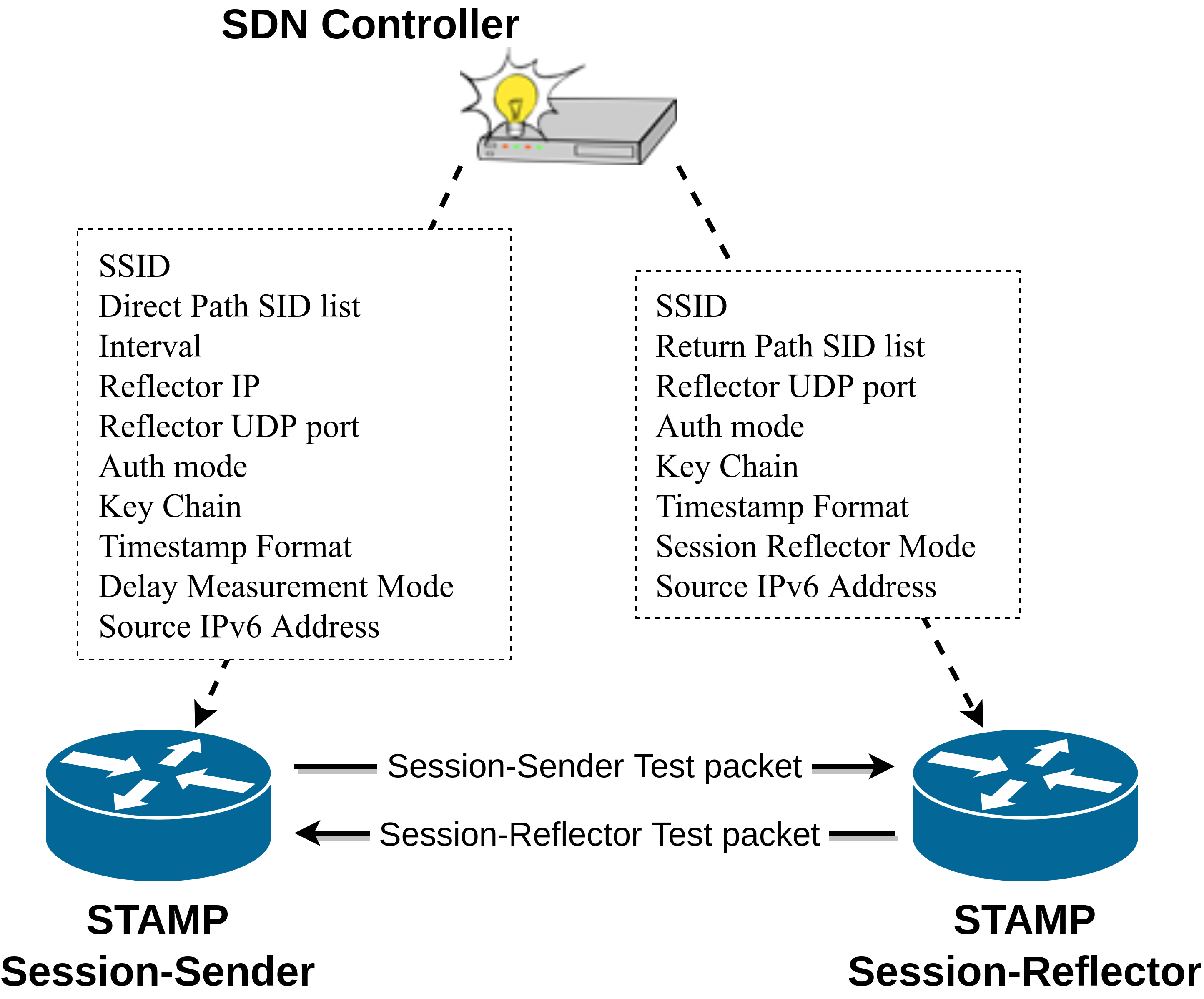}
  \caption{STAMP control protocol.}
  \label{fig:stamp-ctrl-proto}
\end{figure}

We have defined the API offered by the STAMP Session-Sender and by the STAMP Session-Reflector to the SDN controller for the configuration of the STAMP measurement service. The configuration involves setting various parameters, including the STAMP UDP port, the network interfaces on which the STAMP Session-Sender/Session-Reflector expects to receive the STAMP Test packets, and the source IPv6 address to be used in the STAMP Test packets. The controller can also create and manage the STAMP Sessions using the API exposed by the STAMP Session-Sender/Session-Reflector. In particular, to create a STAMP Session, the SDN controller must provide the following parameters: 1) the SSID of the STAMP Session; 2) the SID List of the path under measurement; 3) the interval between two consecutive STAMP Test packets; 4) the source IPv6 address of the STAMP Test packets; 5) the authentication mode (i.e., unauthenticated or authenticated); 6) the timestamp format (i.e., NTP or PTPv2); 7) the delay measurement mode (i.e., one-way or two-way); 8) the IP address of the STAMP Session-Reflector; 9) the STAMP UDP port of the STAMP Session-Sender and Session-Reflector; 10) the Session-Reflector mode (i.e., stateful or stateless). Fig.~\ref{fig:stamp-ctrl-proto} shows the interaction of the SDN controller with the STAMP Session-Sender and Session-Reflector required to create a STAMP Session.

\subsection{Data Collection}
\label{sec:delay-monitoring-framework-data-collection}

The STAMP Session-Sender and STAMP Session-Reflector exchange STAMP Test packets containing the timestamps required to compute the delay. The STAMP Session-Sender collects all the timestamps. The SDN controller can interact with the Session-Sender to fetch the timestamps. In general, there are two approaches the SDN controller can use to fetch the timestamps: \textit{polling mode} and \textit{notification mode}. In polling mode, the controller periodically polls the Session-Sender to gather the collected timestamps. In notification mode, the Session-Sender will \dq{push} the information toward the SDN controller, either by sending the single measurement records or aggregating a set of measurement records in a single notification. In our solution we have implemented the polling mode.

When the measurements records are available to the SDN controller, it can compute the delay of the direct path $d_d$ (i.e., the path from the Session-Sender to the Session-Reflector) and return path $d_r$ (i.e., the path from the Session-Reflector to the Session-Sender):

\begin{equation} \label{eq:delay-direct}
d_{d} = T_2-T_1
\end{equation}
\begin{equation} \label{eq:delay-return}
d_{r} = T_4-T_3
\end{equation}

where $T_1$, $T_2$, $T_3$, and $T_4$ are the four timestamps defined in Section \ref{sec:delay-monitoring-framework-dataplane}, $d_{d}$ and $d_{r}$ are the delay of the direct path and return path, respectively. Of course, the clocks of the Session-Sender and of the Session-Receiver must be synchronized and the accuracy of this delay estimates $d_d$ and $d_r$ depends on the accuracy of the clock synchronization.








%% file: sec/6-stamp_implementations.tex
\section{STAMP for SRv6: Router Implementations}
\label{sec:stamp_implementations}

We have realized an open source prototype of the proposed STAMP for SRv6 solution, see \cite{srv6pm-delay-measurement}. In this section we describe the implementation of the router functionality (Session-Sender and Session-Reflector), in section~\ref{sec:delay_monitoring_everywan} we focus on the SDN Controller and Orchestrator.  

The main Data Plane tasks of the Session-Sender (described in Section \ref{sec:delay-monitoring-framework}) are the following: i) generate and send STAMP Session-Sender Test packets to the STAMP Session-Reflector; ii) receive STAMP Session-Reflector Test packets from the Session-Reflector and collect the timestamps. Concerning the STAMP Session-Reflector, its main Data Plane tasks are: i) receive STAMP Session-Sender Test packets from the Session-Sender; ii) send a STAMP Session-Reflector Test packet to the Session-Sender for each received STAMP Test packet. The Session-Reflector is implemented in its \textit{stateless} version. In the Control Plane, both the Session-Sender and the Session-Reflector interact with the SDN controller by offering an API (see subsection~\ref{sec:stamp_implementations-control-plane}).

As for the Data Plane, we have implemented three versions of the Session-Sender and Session-Reflector with the goal of improving their performance: two User Space implementations (referred to as \textit{basic} and \textit{otimized}, see subsection~\ref{sec:user-space-dataplane-impl}) and a Kernel Space implementation based on the eBPF framework~\cite{ebpf.io}, see subsection~\ref{sec:ebpf-dataplane-impl}. We evaluate and compare the performance of the different implementations in Section~\ref{sec:results}. 


\subsection{Control Plane functionalities}
\label{sec:stamp_implementations-control-plane}

Both the STAMP Session-Sender and Session-Reflector expose a Southbound API that allows an SDN controller to create/start/stop/destroy a STAMP Session and fetch the results of a STAMP Session. This API follows the design ideas discussed in Section \ref{sec:delay-monitoring-framework}. We decided to extend the Southbound API proposed in \cite{ventre2018srv6sdn}, based on the gRPC protocol \cite{grpcio}.
The implementation of our Southbound interface is open-source and available at \cite{srv6pm-delay-measurement}. 
\shortver{The Southbound API supports the following operations: 
\begin{itemize}
    \item \texttt{Init} provides the global configuration parameters (i.e., the parameters common to all the STAMP Sessions) to the STAMP Session-Sender and Session-Reflector.
    \item \texttt{Reset} resets the configuration parameters and stops the packet sniffer.
    \item \texttt{CreateStampSession} prepares the STAMP Session-Sender/Session-Reflector to run a STAMP Session and send/receive the STAMP Test packets. In the Sesssion-Sender, a queue is allocated to store the received measurement results.
    \item \texttt{StartStampSession} and \texttt{StopStampSession} take care of starting and stopping a STAMP Session, respectively. When a session is started in the Session-Sender, a thread is activated that periodically sends STAMP Session-Sender Test packets to the Session-Reflector.
    \item \texttt{DestroyStampSession} removes a STAMP Session and deallocates all the related data structures.
    \item \texttt{GetStampSessionResults} allows the controller to fetch the measurement results (i.e., the timestamps) collected by the STAMP Session-Sender. This RPC is supported only by the Session-Sender as the Session-Reflector does not collect any information during the STAMP Session.
\end{itemize}
A more detailed description of the Southbound interface can be found in \cite{delay-mon-srv6-sdwan-extended}.}


\extended{The Southbound API supports the following operations: i) \texttt{Init}; ii) \texttt{Reset}; iii) \texttt{CreateStampSession}; iv) \texttt{StartStampSession}; v) \texttt{StopStampSession}; vi) \texttt{DestroyStampSession}; vii) \texttt{GetStampSessionResults}.}

\extended{The \texttt{Init} operation is used to provide the global configuration parameters (i.e., the parameters common to all the STAMP Sessions) to the STAMP Session-Sender and Session-Reflector. These parameters include the UDP port of the Session-Sender/Session-Reflector, the interface used to send/receive the STAMP Test packets, and the source IPv6 address of the STAMP Test packets. \texttt{Init} also starts a packet sniffer to intercept all the incoming STAMP Test packets and creates an \textit{ip6tables} rule to drop the STAMP Test packets after they have been processed by STAMP. The STAMP Test packets must be consumed by STAMP and not handled by the kernel. The ip6tables rule prevents STAMP Test packets from being processed by the kernel. The \texttt{Reset} operation resets the configuration parameters and stops the packet sniffer.}

\extended{\texttt{CreateStampSession} prepares the STAMP Session-Sender/Session-Reflector to run a STAMP Session and send/receive the STAMP Test packets. As explained in Section \ref{sec:delay-monitoring-framework}, a STAMP Session is a bidirectional packet exchange between the STAMP Session-Sender and Session-Reflector on a given SRv6 path. The results of a STAMP Session are timestamps that can be used to compute the delay of the SRv6 path under measurement. \texttt{CreateStampSession} allocates all the necessary data structures and provides the configuration parameters specific for the STAMP Session, including the SSID, the Segment List of the path to test, the source IPv6 address of the STAMP Test packets, and the other STAMP parameters described in \cite{rfc8762stamp}. \texttt{StartStampSession} and \texttt{StopStampSession} take care of starting and stopping a STAMP Session, respectively. All the STAMP Sessions are asynchronous. Thus, after starting a STAMP Session, the controller does not need to wait for its completion. Optionally, the controller can also specify the duration of the STAMP Session. Alternatively, the controller can stop the STAMP Session using the \texttt{StopStampSession} RPC. The \texttt{DestroyStampSession} operation removes a STAMP Session, and deallocates all the related data structures.}

\extended{We also implemented a \texttt{GetStampSessionResults} RPC that allows the controller to fetch the measurement results (i.e., the timestamps) collected by the STAMP Session-Sender. This operation is supported only by the Session-Sender as the Session-Reflector does not collect any information during the STAMP Session. Our implementation supports asynchronous fetching. The collected timestamps are stored in the Session-Sender until the controller fetches them. As discussed in Section \ref{sec:delay-monitoring-framework-data-collection}, the controller can compute the delay of the SRv6 path under measurement (see equations \ref{eq:delay-direct} and \ref{eq:delay-return}) based on the collected timestamps.}

\subsection{User Space Implementations for Data Plane}
\label{sec:user-space-dataplane-impl}


In this subsection, we describe our user space implementations of the STAMP Session-Sender and Session-Reflector, compliant with RFC 8762 \cite{rfc8762stamp}, RFC 8972 \cite{rfc8972stampext}, and draft \cite{ietf-spring-stamp-srpm-03}. The implementations are based on the Scapy python library \cite{scapy} and are available as open-source at \cite{srv6pm-delay-measurement}. We have developed a first implementation (referred to as \textit{basic}) and then designed an improved version (referred to as \textit{optimized}. Hereafter, we first describe the \textit{basic} Scapy user space implementation and then we discuss how we have tackled its performance issues with the \textit{optimized} implementation.

The Session-Sender and Session-Reflector leverage the Scapy library to generate the STAMP Test packets. When we started our work, the latest release of Scapy (version 2.4.5) did not implement the RFC 8762 (STAMP). Scapy modular design allows developers to define new protocol layers easily. We have added the support for both STAMP Session-Sender and STAMP Session-Reflector Test packets in unauthenticated mode. Our contribution has been accepted and merged in the mainstream distribution of Scapy, adding the support of the STAMP protocol. Both the Session-Sender Test packet and Session-Reflector Test packet are compliant with the formats defined in RFC 8962 and described in Section \ref{sec:delay-monitoring-framework}. The STAMP Test packets contain the timestamps used to compute the delay. As discussed in Section \ref{sec:delay-monitoring-framework}, STAMP can support two timestamp formats: NTP and PTPv2. Our current implementation only supports NTP timestamps.


After generating the STAMP Test packets, the Session-Sender and the Session-Reflector use the Scapy library to send the packets on the outgoing network interface. In particular, before sending a STAMP Test packet, the Session-Sender adds an UDP header and an IPv6+SRH header to the packet. The UDP header contains the STAMP port of the Session-Reflector as destination port. The SRH contains the Segment List of the path under measurement (i.e., the path from the Session-Sender to the Session-Reflector). The Session-Reflector performs the specular operations adding the proper UDP header and IPv6+SRH header to send the packet to the Session-Sender. Then, the Session-Sender and the Session-Reflector pass the packet to an \texttt{L3RawSocket6}. The \texttt{L3RawSocket6} is a Scapy socket built on top of a AF\_INET6/SOCK\_RAW Linux socket. The Linux kernel adds a Layer 2 header and sends the packet to the destination (i.e., the Session-Reflector or the Session-Sender) according to the usual L2/L3 rules. 

Both the Session-Sender and the Session-Reflector need to process the incoming STAMP Test packets. The Session-Reflector receives the STAMP Session-Sender Test packets from the Session-Sender and it has to reply to these packet by adding the proper timestamps. The STAMP Session-Sender receives STAMP Session-Reflector Test packets from the Session-Reflector and processes them, acting as a measurement data collector. 

The Session-Sender and the Session-Reflector run a dedicated thread to capture, validate and process the STAMP Session Test packets. To capture the incoming STAMP Test packets, the \textit{basic} implementation of Session-Sender uses a Scapy \textit{AsyncSniffer}. The AsyncSniffer captures all the incoming packets received on a given interface and passes the captured packets to a user space callback named \texttt{stamp\_reply\_packet\_received}. This callback drops any non-STAMP Test packet and processes only the valid STAMP Test packets. Since \texttt{stamp\_reply\_packet\_received} operates in user space, calling it for each received packet can have a big impact on the CPU usage. In order to reduce the impact on the CPU usage, it is important to reduce the number of packets processed by the \texttt{stamp\_reply\_packet\_received}. In our implementation, we attach a BPF filter to the AsyncSniffer. This filter allows the AsyncSniffer to capture only the STAMP Test packets by filtering non-STAMP Test packets at kernel level. Thus, \texttt{stamp\_reply\_packet\_received} is invoked only when a STAMP Test packet is received. For each captured STAMP Test packet, \texttt{stamp\_reply\_packet\_received} performs several validation checks. If the packet passes all the validation checks, the Session-Sender extracts the timestamps and collects them in a FIFO queue. The controller periodically can send a \texttt{GetStampSessionResults} command to fetch the latest results from the Session-Sender. The results are kept in the FIFO queue until they are fetched, then they are permanently removed from the queue.

The \textit{basic} implementation of the Session-Reflector performs similar operations to capture STAMP Session-Sender Test packets and send STAMP Session-Reflector Test packets.


During our performance evaluation, we found that the \textit{basic} Scapy solution exhibited very poor performance. 

As explained previously, the \textit{basic} implementation relies on the Scapy AsyncSniffer to capture the STAMP Test packets. AsyncSniffer is implemented using a Linux AF\_PACKET/SOCK\_RAW socket. An AF\_PACKET/SOCK\_RAW socket captures all the packets received on a given interface. The capture process of a plain AF\_PACKET socket is very inefficient, because it uses very limited buffers and requires a system call to capture each packet. 

The second bottleneck of the \textit{basic} implementation is related to the process of building and dissecting the STAMP Test packets. The Session-Sender periodically generates and sends STAMP Test packets to the Session-Reflector. Generating a STAMP Test packet involves several operations, such as building each layer, filling each header with the proper information, stick all the layers together, and computing the checksum. We found that repeating this sequence of operations for building each packet to be transmitted is very expensive. 

Therefore, we designed an improved implementation of the STAMP Session-Sender that mitigates the above described performance issues. We refer to this improved version as \textit{optimized}. This implementation uses the \textit{PACKET\_MMAP} \cite{packetmap} socket option. PACKET\_MMAP improves the capture process by using a circular buffer mapped in user space that can be used to send and receive packets. This buffer is shared between the kernel and our user space application. A shared buffer between the kernel and the user also has the advantage of minimizing packet copies. When a packet arrives, the kernel stores the packet in the buffer. Since the buffer is shared between the kernel and our user space STAMP application, the application can read the packet without issuing any system call.

In order to fix the inefficiencies in the sending procedures, we observed that packets sent in the context of a STAMP Session are very similar to each other. Most of the packet fields are equal for each packet in a STAMP Session. These fields include the SSID, the Segment List, the source and destination IP addresses, and the UDP ports. Few fields need to be changed, such as the timestamp fields and the sequence number contained in the STAMP Test packets. Instead of generating a new packet for each STAMP packet to be sent, the \textit{optimized} implementation of the Session-Sender allocates a STAMP Session-Sender Test packet when the STAMP Session is created (\texttt{CreateStampSession} operation). When a new packet needs to be sent, the Session-Sender only changes the variable fields of the packet (e.g., the timestamps and the sequence number). Then it computes the UDP checksum and sends the packet to the Session-Reflector. In this way, we avoid the overhead related to generating a new STAMP Test packet from scratch. To further improve performance, we save the STAMP Test packet as a bytes array instead of a Python object. In this way, we avoid the overhead due to converting the packet from Python representation to a bytes array before sending it on the network. We also optimized the logic used to parse the received packets. For each received STAMP Session-Reflector Test packet, Scapy performs the so-called \textit{packet dissection}, i.e., it reads the bytes of the packet and builds a Python object to represent the packet. Then, it collects the timestamps from the packet. In the \textit{optimized} solution we bypassed the Scapy dissector and we extract the timestamps directly from the bytes representation of the packets.


As for the Session-Reflector, its \textit{optimized} implementation improves the efficiency of the \textit{basic} version using the same approaches that we have discussed for the the Session-Sender.

The optimized versions of the Session-Sender and Session-Reflector STAMP implementation have been integrated in the EveryWAN prototype as described in Sec.~\ref{sec:delay_monitoring_everywan}.

\subsection{eBPF Implementation for Data Plane}
\label{sec:ebpf-dataplane-impl}

eBPF \cite{ebpf.io} is a Linux technology that enables running programs in kernel space and in a sandboxed environment, without having to deploy ad-hoc kernel modules or change the kernel code. eBPF can offer high performance to specific packet processing tasks. We designed and implemented a proof-of-concept eBPF implementation with the goal to assess its performance.

Our eBPF deployment is based on the HIKe / eCLAT \cite{hike-eclat} \cite{mayer2022ebpf} framework. HIKe (Heal, Improve and desKill eBPF) is a virtual machine abstraction for eBPF. It makes it possible to chain multiple simpler eBPF programs in a larger and more complex program. eCLAT (eBPF Chains Language And Toolset) is a python-like language and programming framework. Its scripts compile to HIKe chains, providing a high-level, simpler language that can be used to compose complex eBPF programs in a modular fashion.


\begin{algorithm}[H]
\caption{HIKe chain high level structure for STAMP Session-Reflector.}
\label{alg:hike_chain}
\begin{algorithmic}
\If{packet is STAMP}
    \State process headers for layers 2, 3, 4
    \State compute UDP checksum
    \State cross connect to layer 2 interface
\Else
    \State pass packet to kernel
\EndIf
\end{algorithmic}
\end{algorithm}

The high-level pseudocode \ref{alg:hike_chain} shows the structure of the HIKe chain for the STAMP Session-Reflector. The chain is attached to the XDP hook on the desired interface and the entire processing is performed without letting the packet enter the Linux kernel networking stack.
The first eBPF program filters only STAMP Test packets, everything else is passed to the kernel without further processing. The chain then manipulates the STAMP fields adding the new timestamps.
Then, the address/port fields in MAC, IPv6 and UDP headers are changed before forwarding the packet. Lastly, the UDP checksum is recalculated and the packet is forwarded on the desired interface.

The Collector implementation is simpler because the packet does not need to be forwarded. The chain comprises a filter so that only STAMP packets are processed, while other packets are sent to the kernel networking stack. Then we have the actual Collector eBPF program. It parses the STAMP payload of the packet and extracts the timestamps. The extracted timestamp records are written inside an eBPF map, accessible from the userspace, so that it is possible to read the measurements.

The code for the eBPF implementation can be found in the repository \cite{hikepkg-stamp}. The deployment and configuration of the eBPF implementation is not integrated in the EveryWAN prototype. The configuration is performed manually as the eBPF proof-of-concept implementation is only used for the performance experiments described in section~\ref{sec:results}. 





%% file: sec/7-delay_monitoring_everywan.tex
\section{Delay Monitoring through EveryWAN Controller}
\label{sec:delay_monitoring_everywan}

\begin{figure}[t!]
  \centering
  \includegraphics[width=\linewidth]{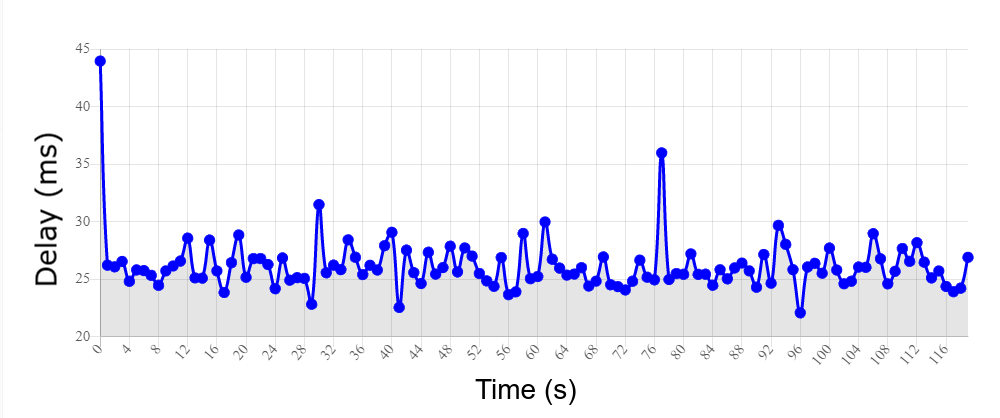}
  \caption{Delay monitoring through the EveryWAN GUI.}
  \label{fig:everywan-gui}
\end{figure}

We integrated the delay monitoring in the EveryWAN prototype. As explained in the EveryWAN white paper \cite{everywan-report}, the EveryEdgeOS controller exposes a Northbound API that allows users to configure the EveryEdge routers and deploy the SD-WAN services. We integrated the STAMP-based delay monitoring capabilities into EveryEdge routers and extended the EveryEdgeOS controller to support STAMP operations. We also extended the Northbound API to offer the basic operations to create, control, and destroy the STAMP Sessions. Furthermore, we added a section to EveryGUI where users can monitor in real time the delay of the deployed SRv6-based VPNs. The result of a measurement session presented on EveryGUI is shown in Figure~\ref{fig:everywan-gui}. In the x-axis there is the time in which each measure is performed. Delays are reported on the y-axis. A walkthrough documentation showing the use of delay monitoring in EveryWAN is available in \cite{srv6-delay-mon}.

In addition to the instant delays, the controller also computes the average delay for both the direct and return paths. The average delay is updated using the \textit{Welford online algorithm} whenever new $d_{d, new}$ and $d_{r, new}$ values are available:


\begin{equation} \label{eq:delay-direct-average}
d_{d, avg} = d_{d, avg} + \frac{d_{d, new} - d_{d, avg}}{N}
\end{equation}
\begin{equation} \label{eq:delay-return-average}
d_{r, avg} = d_{r, avg} + \frac{d_{r, new} - d_{r, avg}}{N}
\end{equation}

where $d_{d, avg}$ is the average delay of the direct path, $d_{r, avg}$ is the average delay of the return path, N is the number of collected delays, and $d_{d, new}$ and $d_{r, new}$ are the new delay values of the direct path and return path, respectively.

%% file: sec/8-monitoring.tex
\section{Experiments and Results}
\label{sec:results}

In this section, we describe the testbed and the methodology used to assess the performance of our STAMP implementations, and we present a comparison between the different implementations.

\subsection{Testbed and Performance Evaluation Methodology}

\begin{figure}
    \centering
    \includegraphics[width=\linewidth]{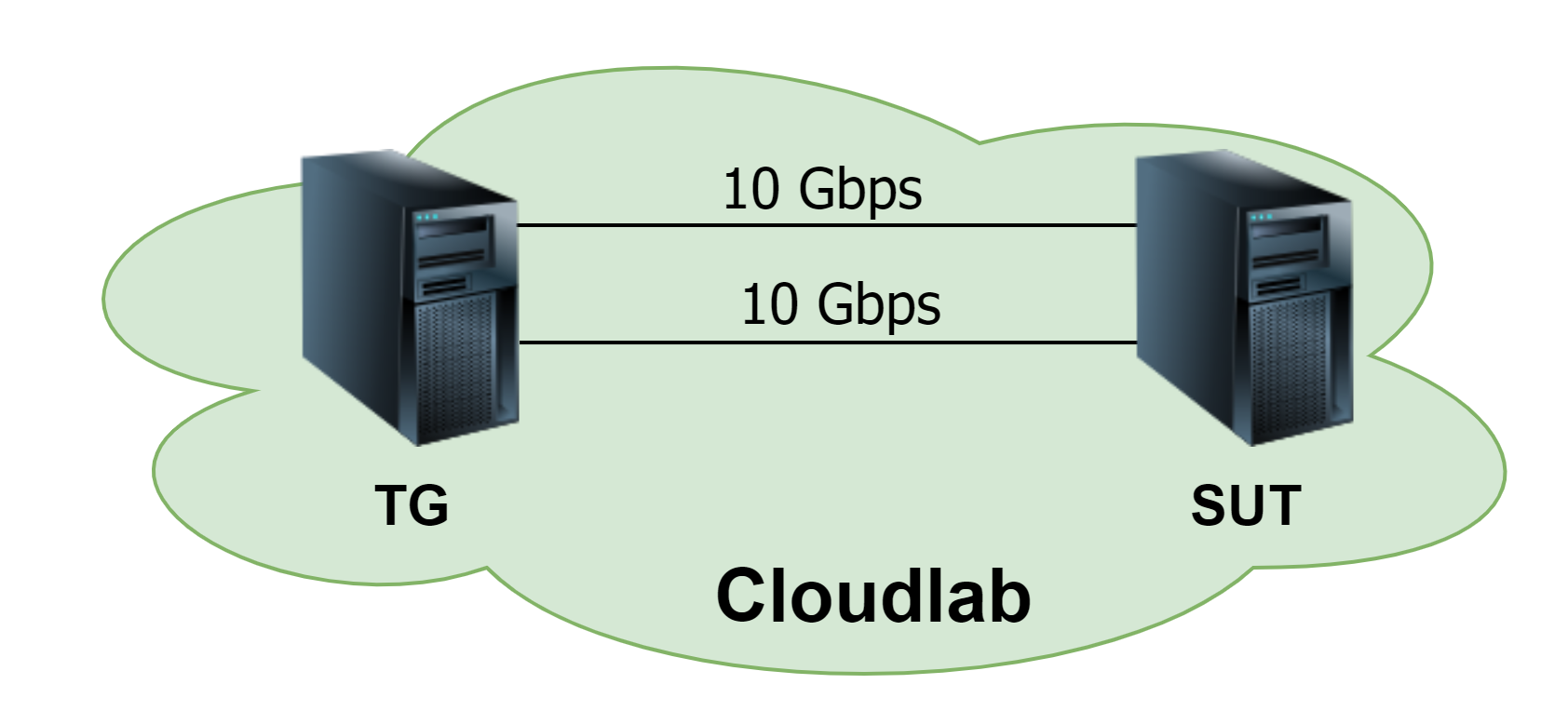}
    \caption{Performance Evaluation Testbed on Cloudlab.}
    \label{fig:cloudlab-testbed}
\end{figure}

To evaluate the performance of our three implementations, we have deployed a testbed according to RFC 2544 \cite{rfc2544}, which provides a methodology for benchmarking network devices. The testbed (shown in Figure \ref{fig:cloudlab-testbed}) includes two nodes: \textit{Traffic Generator (TG)} and \textit{System Under Test (SUT)}. We have deployed our testbed in the Wisconsin cluster of \textit{CloudLab} \cite{cloudlab}, a platform dedicated to scientific research on the future of cloud computing. The testbed nodes (TG and SUT) are bare metal servers equipped with two Intel E5-2630 v3 processors with 16 cores (hyper-threaded) clocked at 2.40GHz, 128 GB of RAM, and two Intel 82599ES 10-Gigabit network interface cards. The TG and SUT nodes are physically connected to the same switch. The two NICs ensure back-to-back connectivity between the two nodes. The logical topology is shown in Fig.~\ref{fig:cloudlab-testbed}. On the TG node, we installed \textit{TRex} \cite{trex}, an open source traffic generator powered by \textit{DPDK} \cite{dpdk}. The SUT node runs an Ubuntu 20.04 LTS Linux distribution with Linux kernel release 5.13 and hosts our STAMP implementations. To control Linux networking capabilities (e.g., network interfaces, routing, and SRv6 behaviors), we installed the 5.13 release of the \textit{iproute2} \cite{iproute2} suite. We also installed \textit{ethtool} 5.13 to configure the hardware capabilities of the NIC, such as offloading \cite{ethtool}. 

To perform the experiments, we used \textit{SRPerf} \cite{abdelsalam2021srperf}, a performance evaluation framework for software and hardware implementations of SRv6. SRPerf orchestrates and automates the execution of the experiments using the TRex Python automation libraries \cite{trex-client-api}. It interacts with the TRex generator installed on the TG. The TG generates packets using the TRex traffic generator and sends them to the SUT. The SUT processes the received packets. The TG evaluates the maximum throughput that can be processed by the SUT. SRPerf supports different throughput measurements, such as \textit{No-Drop Rate (NDR)}, \textit{Partial Drop Rate (PDR)}, and \textit{Maximum Receive Rate (MRR)}. In our experiments, we used the Partial Drop Rate at a 0.5\% drop ratio (in short, PDR@0.5\%) as throughput measurement, which is defined as the maximum packet rate at which the packet drop ratio is less than or equal to 0.5\%. For further details on this metric and how it is evaluated by the SRPerf tools, we refer to \cite{abdelsalam2021srperf}.

Our goal is to evaluate the impact of STAMP measurement procedures on the packet processing capabilities of a Linux software router. As a reference, we consider the scenario in which the router is only processing regular data packets, then we intermix regular data packets with STAMP measurement packets in different percentages.

For the processing of regular data packets, we consider an SRv6 ingress node that performs packet encapsulation: it receives IPv6 packets and applies the H.Encaps behavior to encapsulate the packets in an outer IPv6+SRH packet. Therefore, in our baseline scenario the TG generates IPv6 packets, the SUT receives the packets on one interface, performs the encapsulation, and forwards the packets on the second interface. 

For the processing of the STAMP measurement packets, we have considered two cases:

\begin{enumerate}
    \item the SUT is configured as a STAMP Session-Reflector, it receives STAMP Session-Sender Test packets, processes them, and for each STAMP Test packet it sends a STAMP Session-Reflector Test packet to the TG;
    \item the SUT is configured as a STAMP Session-Sender, it receives STAMP Session-Reflector Test packets, extracts, and collects the timestamps from the packets, performing the role of the \emph{Collector}.
\end{enumerate}

The impact of STAMP measurements is evaluated by changing the fraction of STAMP packets and measuring the packet processing capacity using the PDR@0.5\% metric. When the SUT acts as a Session-Reflector (case 1), the methodology to evaluate the packet drop ratio described above can be applied easily, as both the data packets and the STAMP test packets are forwarded back by the SUT towards the TG (the data packets are encapsulated, the STAMP packets are processed and properly updated). To evaluate the packet drop ratio, the TG simply compares the number of transmitted and received packets in an experiment session (summing up the data and STAMP test packets). On the other hand, when the SUT acts as a Session-Sender/Collector (case 2), it does not forward the received STAMP test packets back to the TG, because it receives the STAMP packets and produces the measurement records. Therefore, the TG cannot simply count the packets transmitted back by the SUT to evaluate the packet drop ratio. In fact, the number of packets correctly processed by the SUT corresponds to the sum data packets that are forwarded back and the STAMP test packets that are properly processed by the SUT (i.e., by collecting the STAMP measurement metrics). A STAMP packet that is not processed by the SUT must count as a dropped packet. Therefore, the TG must retrieve the counter of processed STAMP packets from the router under test after each experiment session. To solve this problem, we have designed and implemented a gRPC based API. The SUT/router acts as a gRPC server, whereas a gRPC client in the TG queries the server after each experiment session and retrieves the number of processed STAMP packets. In this way, the TG can sum up this number with the number of received data packets and can properly evaluate the packet drop ratio. 

To run the performance experiments, a careful configuration of the SUT node is needed because we need to saturate the capacity of a CPU to measure the PDR@0.5\% metric. Therefore, we need that all tasks of our interest are executed by the selected CPU and we need to avoid that any other task is executed in the same CPU. \shortver{A detailed discussion on these aspects can be found in the Appendix of \cite{delay-mon-srv6-sdwan-extended}.} 
\extended{A detailed discussion on these aspects can be found in the Appendix.} A walk-through documentation of how to setup the testbed and run the experiment is available in \cite{srv6-delay-mon}.

\subsection{Performance analysis}

We report several experiments to evaluate the impact of our Session-Sender and Session-Reflector implementations on the user traffic. First, we evaluate the forwarding capability in the scenario with only data traffic (no STAMP test packets) without running any STAMP implementation. We consider this throughput as our baseline. Then, we run the Session-Sender or the Session-Reflector on the SUT and we evaluate the maximum achievable throughput for different combinations of data and STAMP test packets using our three different STAMP implementations.

\begin{figure}
    \centering
    \includegraphics[width=0.8\linewidth]{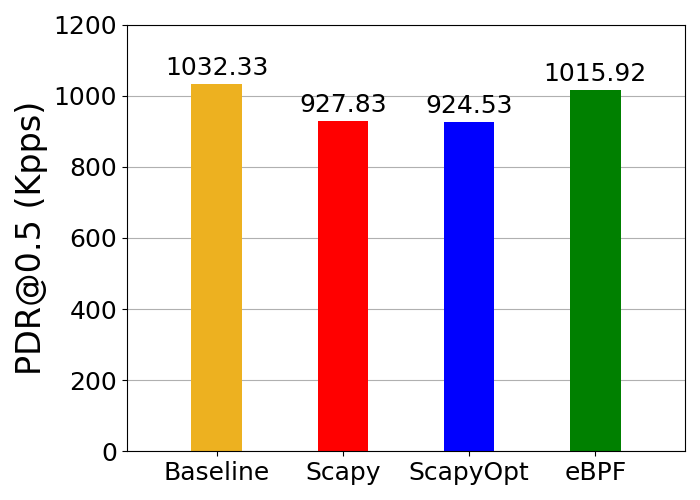}
    \caption{Collector throughput, only data traffic.}
    \label{fig:0stamp-collector-barplot}
\end{figure}

\begin{figure}
    \centering
    \includegraphics[width=0.8\linewidth]{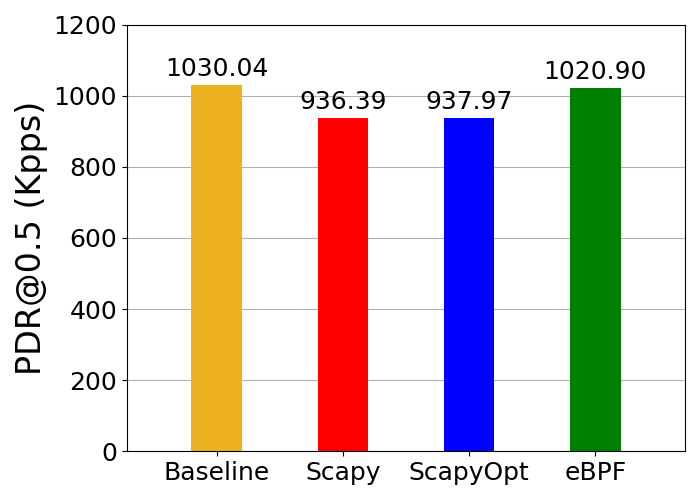}
    \caption{Reflector throughput, only data traffic.}
    \label{fig:0stamp-reflector-barplot}
\end{figure}

\begin{figure}[t]
  \centering
  \begin{subfigure}[htbp]{0.48\linewidth}
    \centering
    \includegraphics[width=\textwidth]{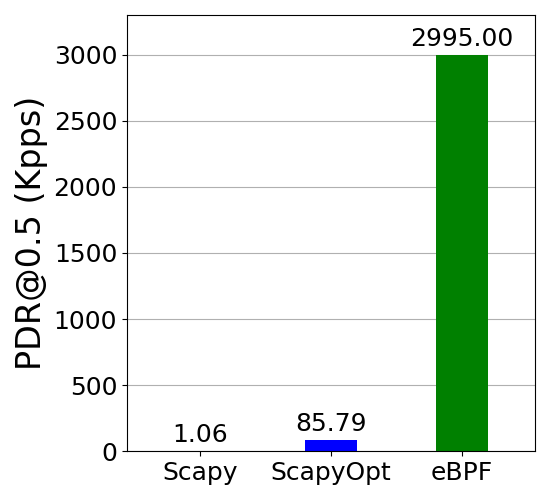}
    \caption{Collector.}
    \label{fig:100stamp-collector-barplot}
  \end{subfigure}
  \begin{subfigure}[htbp]{0.48\linewidth}
    \centering
    \includegraphics[width=\textwidth]{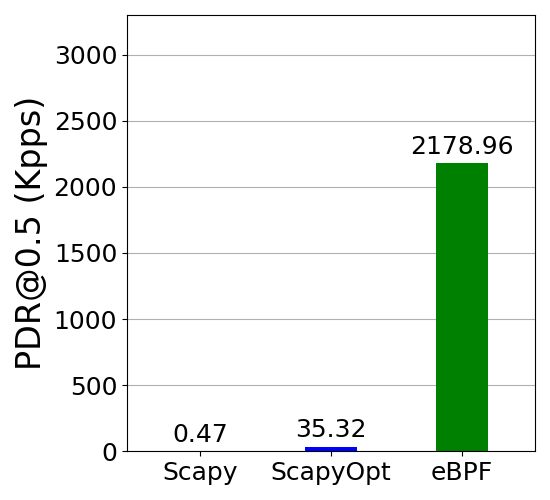}
    \caption{Reflector.}
    \label{fig:100stamp-reflector-barplot}
  \end{subfigure}\hspace{2pt}
  \caption{Throughput, only STAMP traffic.}
  \label{fig:100stamp-barplot}
\end{figure}

\begin{figure}
    \centering
    \includegraphics[width=\linewidth]{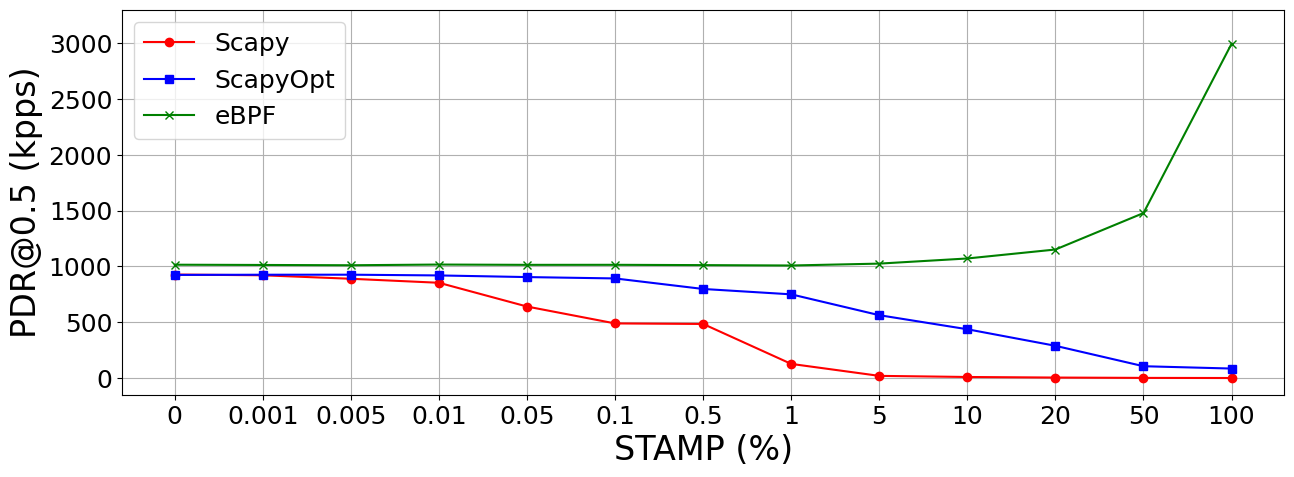}
    \caption{Collector throughput, only STAMP traffic.}
    \label{fig:stamp-collector-encap}
\end{figure}

\begin{figure}
    \centering
    \includegraphics[width=\linewidth]{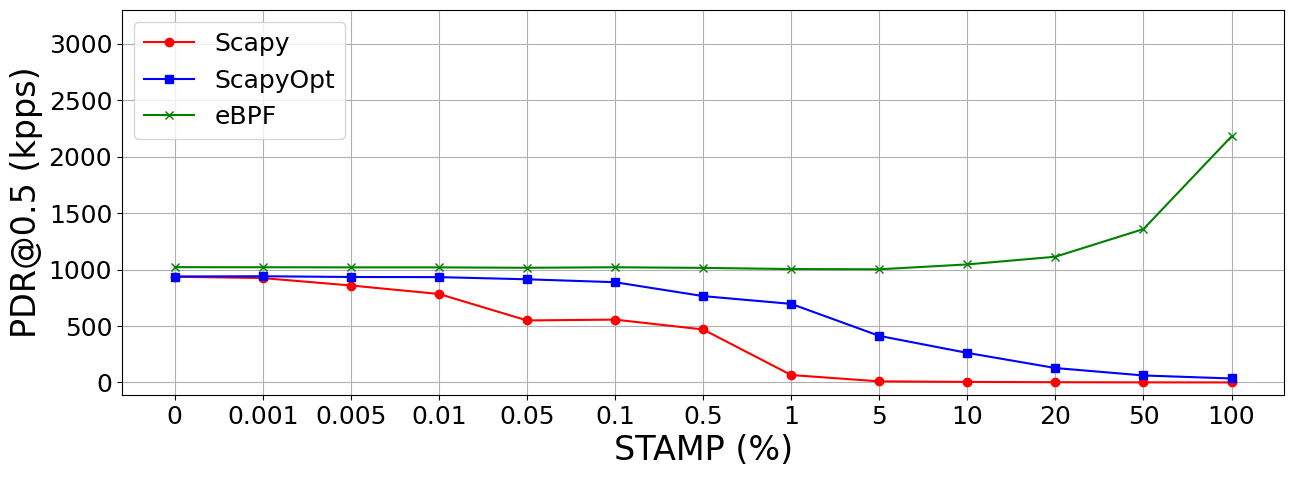}
    \caption{Reflector throughput, only STAMP traffic.}
    \label{fig:stamp-reflector-encap}
\end{figure}

The forwarding capacity of the node is measured using the PDR@0.5\% metric as discussed in the previous subsection. The results reported in Figs.~\ref{fig:0stamp-collector-barplot}-\ref{fig:stamp-reflector-encap} are always the average of 10 evaluations (every single evaluation is carried out using the SRPerf tool \cite{abdelsalam2021srperf}). We do not report error bars with confidence intervals in our figures, as we obtained stable results and the 95\% confidence intervals are so close to the average that they are not noticeable. \shortver{The tables with the detailed results are reported in the Appendix of \cite{delay-mon-srv6-sdwan-extended}.} 
\extended{The tables with the detailed results are reported in the Appendix.}

The comparison among the STAMP Session-Sender/Collector implementations is shown in Fig.~\ref{fig:0stamp-collector-barplot}. The Scapy implementations suffer a 10.4\% performance degradation compared to the baseline performance. This performance degradation is due to the fact that even if there are no STAMP Test packets to be processed, the Session-Sender still has to look at all the incoming packets to capture the STAMP Test packets. This operation is very efficient, as it is executed in kernel mode. Both user space implementations have the same performance ($\approx$925 kpps). The reason lies in the fact that even if the two implementations differ greatly in the processing of STAMP Test packets, the mechanisms used to filter the STAMP Test packets are the same. Thus, when there is only data traffic, the two implementations exhibit the same performance degradation. The packet rate of the \textit{eBPF-based} implementation ($\approx$1016 kpps) is higher than the two user space implementations. This is due to the fact that the HIKe eBPF chain contains a more efficient eBPF filter with respect to the filter of the user space implementation. Since this test is performed without STAMP packets, the performance is only affected by the filter that the packet traverses before being sent to the kernel networking stack. The performance drop of the \textit{eBPF-based} implementation with respect to the baseline is 1.6\%.

The STAMP Session-Reflector implementations exhibit the same behavior when processing only data traffic. A comparison among the Session-Reflector implementations is shown in Fig.~\ref{fig:0stamp-reflector-barplot}.

We evaluated the PDR@0.5\% in the opposite scenario in which there is only measurement traffic (i.e., only STAMP Test packets). The results are shown in Fig.~\ref{fig:100stamp-barplot}.

Regarding the Session-Reflector (shown in Fig.~\ref{fig:100stamp-collector-barplot}), the \textit{basic} implementation reaches a packet rate of $\approx$1.06 kpps, which is much lower than the other two implementations. As discussed in Section \ref{sec:stamp_implementations}, the reasons for this poor performance are related to the inefficiency of the Scapy AsyncSniffer and the high overhead of the Scapy builder and dissector. In the \textit{optimized} implementation, we mitigated these issues. This allows the Session-Sender to reach an higher packet rate, $\approx$85.8 kpps. The performance of \textit{eBPF-based} implementation is much higher ($\approx$2995 kpps). The reason is that eBPF performs all the processing in kernel space, while \textit{optimized} is a user space solution.

Concerning the performance of the Session-Reflector (shown in Fig.~\ref{fig:100stamp-reflector-barplot}), we observe the same trend (Fig.~\ref{fig:100stamp-reflector-barplot}). The \textit{basic} implementation reaches a packet rate of $\approx$470 pps, which is lower than the packet rates of the \textit{optimized} ($\approx$35.3 kpps) and \textit{eBPF-based} implementation ($\approx$2179 kpps). The performance of the Session-Sender is always better than the Session-Reflector. The reason is that the Session-Sender processing is less expensive than the Session-Reflector processing. For each received STAMP Session-Reflector packet, the Session-Sender must collect and store the timestamps. Instead, when the Session-Reflector receives a STAMP Session-Sender Test packet, it must generate a STAMP Session-Reflector Test packet and forward the packet towards the Session-Sender. These operations are much more expensive than storing the timestamps.

Clearly, the scenario described above with only measurement traffic is unrealistic. We only use it to assess and compare the performance of the different implementations. In real scenarios, the measurement traffic (i.e., STAMP) is a small fraction of the overall traffic and will never reach 100\% link capacity. For this reason, we analysed the performance considering different fraction of STAMP measurement packets.

Fig.~\ref{fig:stamp-collector-encap} shows the maximum achievable throughput for the Session-Sender, varying the fraction of STAMP measurement packets. The \textit{basic} implementation starts at $\approx$927.8 kpps at 0\% STAMP, drops to $\approx$641.3 kpps (at 0.05\% STAMP) and $\approx$20.3 kpps (at 5\% STAMP), and then it continues to slowly drop to $\approx$1.06 kpps (100\% STAMP). The throughput of the \textit{optimized} implementation starts at $\approx$924.5 kpps and remains stable until the measurement traffic is 0.1\% of the total traffic. The packet rate of the \textit{eBPF-based} implementation starts at $\approx$1015.9 kpps when there is no measurement traffic (i.e., no STAMP packets) and it remains almost stable until the measurement traffic is 10\% of the total traffic. Then, we observe a trend in contrast with the two user space implementations. The performance goes up to $\approx$1152.1 kpps when the measurement traffic is 20\% of the total traffic and reaches $\approx$2994.9 kpps when the measurement traffic is 100\%.

The reason why the eBPF implementation starts with a higher throughput (PDR@0.5\%) when the STAMP traffic is low, is that its BPF filter used to select the STAMP traffic is lighter than the one used by the Scapy implementations. When the percentage of STAMP traffic is very low, it does not affect the overall performance and the filtering is the only factor that plays a role. When the STAMP traffic increases, the throughput of the eBPF implementation increases because the STAMP packets are not sent to the kernel networking stack and they are processed faster by our eBPF program than the SRv6 packets that the kernel is encapsulating. On the other hand, the Scapy implementations process the STAMP packets in the user space, hence the performance is reduced when the fraction of STAMP packets increases.

The Session-Reflector throughput for different value of the percentage of STAMP measurement packets is shown in Fig.~\ref{fig:stamp-reflector-encap}. The results for the three implementations are consistent with what we have discussed for the Session-Sender/Collector implementation. For high value of the percentage of STAMP traffic, it can be noted that the performance is slightly lower, this is because the Session-Reflector sends back the STAMP measurement packets. Apparently, this is heavier than storing the STAMP measurement records as done by the Session-Sender/Collector. 

%% file: sec/9-related_works.tex
\section{Related Works}
\label{sec:related-works}

Several solutions have been proposed for performance monitoring in a network. Some of them like Nagios \cite{nagios} and Zabbix \cite{zabbix} focus on the monitoring of network devices. Other solutions like Ceilometer \cite{ceilometer} target cloud environments. Concerning SDN, several solutions have been proposed. OpenNetMon \cite{adrichem2014opennetmon} is a framework to measure throughput, delay, and packet loss in OpenFlow networks. A monitoring framework for SDN Virtual Networks is proposed in \cite{yang2020monitoring}. Other solutions for OpenFlow networks can be found in \cite{queiroz2019approach} and \cite{santos2015network}. \cite{tsai2018network} proposes a review of the monitoring techniques used in SDN.

IETF worked on the standardization of a protocol to measure the performance of IP and MPLS networks. This protocol is defined in RFC 4656 \cite{rfc4656} and it is called \textit{One-Way Active Measurement Protocol
(OWAMP)}. OWAMP only focused on the one-way performance metrics, such as one-way delay and one-way packet loss. Another protocol was defined later, called \textit{Two-Way
Active Measurement Protocol (TWAMP)}. TWAMP (defined in RFC 5357 \cite{rfc5357}) introduced the two-way measurements. RFC 5357 defines both the test protocol (i.e., the format of the messages exchanged to collect the measures) and the control protocol (i.e., the protocol used to setup the parameters required by the measurement session). RFC 8762 \cite{rfc8762stamp} introduces a new protocol, known as \textit{Simple Two-Way Active Measurement Protocol (STAMP)}. RFC 8972 \cite{rfc8972stampext} proposes optional extensions, such as TLV (Type-Length-Value) coding to specify the
Return Path. Later on, the STAMP protocol has been extended to support SR networks (both SR-MPLS and SRv6) \cite{ietf-spring-stamp-srpm-03}. This solution can measure metrics like delay or packet loss of a SRv6 path. The measurement mechanism is based on packets exchanged on the SRv6 path under measurement. These packets carry information used to compute the performance.

In \cite{loreti2020pfplm}, the authors described a per-flow packet loss measurement solution based on the \textit{alternate marking} method called PF-PLM. They also proposed and compared two different implementations of the proposed solution, realized extending Netfilter/Xtables and IP set Linux frameworks, respectively. In our previous work \cite{loreti2021srv6pm}, we proposed an open source solution for Performance Monitoring of SRv6 networks, called SRv6-PM. SRv6-PM includes a cloud-native infrastructure that supports ingestion, processing, storage and visualization of PM data. We also provided an implementation based on the eBPF framework. Both works focused on packet loss monitoring.

In \cite{zhao2022sra}, the authors described SRA, a user space implementation of the SRv6 data plane based on AF XDP. The proposed solution supports a custom SRv6 behavior called End.DM which enables the measurement of the delay in SRv6 networks. SRA collects the timestamps in each node of the SRv6 path. Our solution does not implement an SRv6 dataplane, it only implements the STAMP protocol and leaves the SRv6 packets to the Linux kernel. Moreover, STAMP is focused on the end-to-end delay, so it is not needed to record all the intermediate nodes timestamps.


%% file: sec/10-conclusions.tex
\section{Conclusion}
\label{sec:conclusion}

In this paper, we proposed a solution to support the delay monitoring of SRv6 SD-WAN services. Our solution is based on the STAMP protocol and its extensions to support performance measurements in SRv6 networks, currently under discussion in the IETF. The main components of the solution are the STAMP Session-Sender and Session-Reflector which run in the SRv6 routers and perform the delay monitoring operations in the data plane. These data plane components need to be configured to execute the monitoring procedures. We defined and implemented an API that allows an SDN controller to interact with the Session-Sender and Session-Reflector.  We integrated the proposed solution in EveryWAN, an SD-WAN open source prototype. Therefore, we deployed and tested a complete open source framework for delay monitoring of SRv6 based SD-WANs. In this respect, we have given a positive answer to the first two research and technological questions outlined in the introduction: i) the proposed approach based on IETF standards and current Internet drafts is an effective solution for delay monitoring of SRv6 networks; ii) we were able to implement the Delay Monitoring in an open source prototype based on Linux software routers, covering both the data plane aspects and the control plane aspects.

Then, we have addressed the research questions related to the performance impact of delay monitoring procedures on a Linux software router. We have implemented the proposed solution in three different versions and executed a number of performance experiments to evaluate and compare the three implementations. We have started with a naive user space implementation of STAMP based delay monitoring, but we realized that its performance was poor, with a high reduction of the forwarding capacity of the software router. We have optimized the user space implementation, achieving an acceptable performance impact. In particular, with the optimized user space implementation the impact is acceptable when the fraction of measurement packets is kept within reasonable limits (e.g. less than 0.1\%). We think that these limits will not be exceeded under practical operational conditions, as the number of measurement packets will always be a small fraction of the data traffic. Therefore, we have integrated the optimized user space implementation in our open source SD-WAN framework, which now offers a running prototype of the delay monitoring solution. We further considered a third implementation, based on the Linux eBPF technology. This proof-of-concept implementation providee a positive answer to question about the feasibility of delay monitoring in SD-WANs with negligible impact on the forwarding capability of a Linux software router.

%% file: sec/11-appendices.tex
\appendices
\section{Additional details on the experiments}

\subsection{Configuration of the SUT node}

We configured the SUT node according to the guidelines described in \cite{abdelsalam2021srperf}. In particular, we disabled the hyper-threading feature of the SUT node using the \textit{sysctl} Linux utility. We used the SMP IRQ affinity features to bind all the NIC receive queues to a single CPU core. This prevents the kernel from distributing the received packets across multiple CPU cores. In this way, we force all the incoming traffic (i.e., STAMP traffic and user traffic) to be processed by a single CPU core. For the user space implementation, we used the \textit{taskset} utility to bind the user space process to the same CPU. In order to make the experiments independent of the NIC hardware capabilities, we also disabled all the NIC hardware offloading capabilities, including checksum offload, Large Receive Offload (LRO), Generic Receive Offload (GRO), and Generic Segmentation Offload (GSO).

\subsection{eBPF Implementation of the Session-Sender/Collector}

To evaluate the performance of the Session-Sender/Collector, we need to count the number of correctly processed STAMP test packets during a test session. 

For testing the eBPF implementation, it is not possible to count this number after the test session, because the records that are written in the eBPF maps exceed the capacity of the map, so they are overwritten. Therefore, we have developed a modified version of the eBPF implementation of the Session-Sender/Collector, which counts the correctly processed packets and writes them in a map record. This counting operation can impact performance because it requires additional processing per each received packet. Therefore, we modified the implementation so that the same number of operations is executed by the regular implementation and by the version that does the counting. This obviously means that the version that does the counting is not actually storing the measurement records, but this is fine for our purposes. 

The regular implementation performs a read operation and a write operation per each received packet, using these two maps:
\begin{itemize}
    \item A map containing a single timestamp (64 bits) contains data used to reconstruct the local timestamp upon receiving the packet. This is accessed with a \textbf{read} operation for each received packet.
    \item A map containing 4 timestamps (64 bits each) is used to store the STAMP data record after the packet is received. This is accessed with a \textbf{write} operation for each received packet.
\end{itemize}

The modified implementation for counting purposes performs one read operation to read the counter with the number of packets processed and a write operation to write the updated counter. To keep the complexity identical to the regular implementation, we would need to read a 64 bit counter from the first map, and write the incremented counter to a map with 4*64 bits values, which of course is not possible. We carried out the test twice, the first time reading and writing to the map with 64 bits values, and the second time using the map with 4*64 bits values. With the first test, we get an overestimation of the processing capability, because we are performing a shorter write operation (64 bits instead of 4*64 bits). With the second test we get an underestimation of the processing capabilities, because we are performing a longer read operation (4*64 bits instead of 64 bits). In our results shown in Fig.~\ref{fig:stamp-collector-encap} we report the average between the two tests. The difference between the two tests is only noticeable in the two experiments with a high percentages of STAMP traffic (50\% and 100\% in Fig.~\ref{fig:stamp-collector-encap}).  In particular, the PDR@0.5\% metric at 100\% STAMP traffic is $3.3 Mpps$ for the first test and $2.6 Mpps$ for the second test. The actual achievable throughput is in between these values, but we are not interested in its precise evaluation, because the interesting part of the Fig.~\ref{fig:stamp-collector-encap} is in the left part, where the percentage of STAMP measurement packets is below 1\%.

\subsection{Implementation of gRPC server}

In order to count the STAMP packets processed by the Collector, we implemented a gRPC API GetResultsCounter() that allows the TG to retrieve the number of STAMP Test packets received and processed by the SUT. In our first implementation, we run the gRPC server on a dedicated thread running on the same CPU as the Collector. During our experiments, we discovered that the gRPC server impacted on the packet processing capability of the router, which resulted in inaccurate measurements. To fix this issue, we had to move the gRPC server to a different process and bind the process to a different CPU.

\subsection{Results}

In the following tables, we report the details of the experiment results reported in Figs.~\ref{fig:0stamp-collector-barplot}-\ref{fig:stamp-reflector-encap}.

Each value of PDR@0.5\% is averaged over 10 evaluations performed with the SRPerf tool \cite{abdelsalam2021srperf} over the testbed in Cloudlab shown in Fig.\ref{fig:cloudlab-testbed}. We report the average (Avg) and standard deviation (SD) in kpps ($10^3$ packet/s) and the Coefficient of Variation (CV), i.e. the ratio between the Standard Deviation and the average. We evaluate the 95\% confidence interval, denoted as CI95, and report in the table the ratio between CI95 and the average. As we can see, the ratio between CI95 and the average is always very low, for this reason we have not plotted error bars with confidence intervals in Figs~\ref{fig:0stamp-collector-barplot}-\ref{fig:stamp-reflector-encap}. Note that in Table \ref{tab:basic-reflector} the values for SD, CV and CI95 are zero for high percentages of STAMP traffic. This happens because the throughputs are very low and the resolution of SRPerf is also low, with the consequence that all the measures are equal. This is not a problem because we are not interested in measuring with high precision the throughputs in the order of few kpps. We are simply assessing that the performance is very poor (not acceptable) when the maximum rate of STAMP test packets that can be processed is in the order of one thousand packets per second.

\begin{center}
\begin{table*}
\begin{tabular}{ | m{4.5em} | m{1cm}| m{2.7em} | m{2.9em} | m{2.9em} | m{2.7em} | m{2.7em} | m{2.7em} | m{2.7em} | m{2.7em} | m{2.7em} | m{2.7em} | m{2.7em} | m{2.7em} | m{2.5em} | } 
\toprule
& \textbf{Baseline} & \textbf{0\%} & \textbf{0.001\%} & \textbf{0.005\%} & \textbf{0.01\%} & \textbf{0.05\%} & \textbf{0.1\%} & \textbf{0.5\%} & \textbf{1\%} & \textbf{5\%} & \textbf{10\%} & \textbf{20\%} & \textbf{50\%} & \textbf{100\%} \\
\midrule
  \hline
  Avg [kpps] & 1030.04 & 936.39 & 924.31 & 858.48 & 783.61 & 549.15 & 556.49 & 469.25 & 65.95 & 9.95 & 4.92 & 2.44 & 0.94 & 0.47 \\ 
  \hline
  SD [kpps] & 4.969 & 5.087 & 1.703 & 0.692 & 1.899 & 13.823 & 12.003 & 16.776 & 0.089 & 0.024 & 0.000 & 0.000 & 0.000 & 0.000 \\ 
  \hline
  CV & 0.48\% & 0.54\% & 0.18\% & 0.08\% & 0.24\% & 2.52\% & 2.16\% & 3.57\% & 0.13\% & 0.24\% & 0.00\% & 0.00\% & 0.00\% & 0.00\% \\ 
  \hline
  CI$_{95}$ / Avg & 0.29\% & 0.33\% & 0.11\% & 0.05\% & 0.15\% & 1.51\% & 1.29\% & 2.14\% & 0.08\% & 0.15\% & 0.00\% & 0.00\% & 0.00\% & 0.00\% \\ 
  \hline
\end{tabular}
\caption{Scapy Basic Reflector.}
\label{tab:basic-reflector}
\end{table*}
\end{center}

\begin{center}
\begin{table*}
\begin{tabular}{ | m{4.5em} | m{1cm}| m{2.7em} | m{2.9em} | m{2.9em} | m{2.7em} | m{2.7em} | m{2.7em} | m{2.7em} | m{2.7em} | m{2.7em} | m{2.7em} | m{2.7em} | m{2.7em} | m{2.5em} | } 
\toprule
& \textbf{Baseline} & \textbf{0\%} & \textbf{0.001\%} & \textbf{0.005\%} & \textbf{0.01\%} & \textbf{0.05\%} & \textbf{0.1\%} & \textbf{0.5\%} & \textbf{1\%} & \textbf{5\%} & \textbf{10\%} & \textbf{20\%} & \textbf{50\%} & \textbf{100\%} \\
\midrule
  \hline
  Avg [kpps] & 1032.85 & 937.97 & 940.42 & 933.79 & 932.69 & 913.55 & 887.65 & 764.59 & 695.74 & 412.44 & 261.78 & 127.81 & 62.02 & 35.32 \\ 
  \hline
  SD [kpps] & 2.021 & 3.081 & 1.930 & 3.540 & 2.134 & 3.064 & 2.518 & 1.834 & 1.929 & 1.801 & 0.839 & 0.157 & 0.136 & 0.091 \\ 
  \hline
  CV & 0.20\% & 0.33\% & 0.21\% & 0.38\% & 0.23\% & 0.34\% & 0.28\% & 0.24\% & 0.28\% & 0.44\% & 0.32\% & 0.12\% & 0.22\% & 0.26\% \\ 
  \hline
  CI$_{95}$ / Avg & 0.12\% & 0.20\% & 0.12\% & 0.23\% & 0.14\% & 0.20\% & 0.17\% & 0.14\% & 0.17\% & 0.26\% & 0.19\% & 0.07\% & 0.13\% & 0.16\% \\ 
  \hline
\end{tabular}
\caption{Scapy Opt Reflector.}
\end{table*}
\end{center}

\begin{center}
\begin{table*}
\begin{tabular}{ | m{4.5em} | m{1cm}| m{2.7em} | m{2.9em} | m{2.9em} | m{2.7em} | m{2.7em} | m{2.7em} | m{2.7em} | m{2.7em} | m{2.7em} | m{2.7em} | m{2.7em} | m{2.7em} | m{2.5em} | } 
\toprule
& \textbf{Baseline} & \textbf{0\%} & \textbf{0.001\%} & \textbf{0.005\%} & \textbf{0.01\%} & \textbf{0.05\%} & \textbf{0.1\%} & \textbf{0.5\%} & \textbf{1\%} & \textbf{5\%} & \textbf{10\%} & \textbf{20\%} & \textbf{50\%} & \textbf{100\%} \\
\midrule
  \hline
  Avg [kpps] & 1028.23 & 1020.90 & 1020.03 & 1019.01 & 1018.91 & 1015.34 & 1019.98 & 1014.47 & 1004.06 & 1001.94 & 1045.24 & 1113.36 & 1358.18 & 2178.96 \\ 
  \hline
  SD [kpps] & 3.137 & 2.705 & 4.696 & 2.463 & 3.556 & 2.585 & 1.019 & 1.015 & 2.884 & 1.114 & 8.388 & 4.662 & 4.659 & 2.814 \\ 
  \hline
  CV & 0.31\% & 0.26\% & 0.46\% & 0.24\% & 0.35\% & 0.25\% & 0.10\% & 0.10\% & 0.29\% & 0.11\% & 0.80\% & 0.42\% & 0.34\% & 0.13\% \\ 
  \hline
  CI$_{95}$ / Avg & 0.18\% & 0.16\% & 0.28\% & 0.15\% & 0.21\% & 0.15\% & 0.06\% & 0.06\% & 0.17\% & 0.07\% & 0.48\% & 0.25\% & 0.21\% & 0.08\% \\ 
  \hline
\end{tabular}
\caption{eBPF Reflector.}
\end{table*}
\end{center}

\begin{center}
\begin{table*}
\begin{tabular}{ | m{4.5em} | m{1cm}| m{2.7em} | m{2.9em} | m{2.9em} | m{2.7em} | m{2.7em} | m{2.7em} | m{2.7em} | m{2.7em} | m{2.7em} | m{2.7em} | m{2.7em} | m{2.7em} | m{2.5em} | } 
\toprule
& \textbf{Baseline} & \textbf{0\%} & \textbf{0.001\%} & \textbf{0.005\%} & \textbf{0.01\%} & \textbf{0.05\%} & \textbf{0.1\%} & \textbf{0.5\%} & \textbf{1\%} & \textbf{5\%} & \textbf{10\%} & \textbf{20\%} & \textbf{50\%} & \textbf{100\%} \\
\midrule
  \hline
  Avg [kpps] & 1032.33 & 927.83 & 921.26 & 890.05 & 854.10 & 641.28 & 489.97 & 485.22 & 127.43 & 20.28 & 10.28 & 5.17 & 2.10 & 1.06 \\ 
  \hline
  SD [kpps] & 3.735 & 2.424 & 2.526 & 2.220 & 1.588 & 2.023 & 0.761 & 11.880 & 0.772 & 0.274 & 0.179 & 0.223 & 0.063 & 0.039 \\ 
  \hline
  CV & 0.36\% & 0.26\% & 0.27\% & 0.25\% & 0.19\% & 0.32\% & 0.16\% & 2.45\% & 0.61\% & 1.35\% & 1.74\% & 4.31\% & 2.98\% & 3.71\% \\ 
  \hline
  CI$_{95}$ / Avg & 0.22\% & 0.16\% & 0.16\% & 0.15\% & 0.11\% & 0.19\% & 0.09\% & 1.47\% & 0.36\% & 0.81\% & 1.04\% & 2.59\% & 1.79\% & 2.23\% \\ 
  \hline
\end{tabular}
\caption{Scapy Basic Collector.}
\end{table*}
\end{center}

\begin{center}
\begin{table*}
\begin{tabular}{ | m{4.5em} | m{1cm}| m{2.7em} | m{2.9em} | m{2.9em} | m{2.7em} | m{2.7em} | m{2.7em} | m{2.7em} | m{2.7em} | m{2.7em} | m{2.7em} | m{2.7em} | m{2.7em} | m{2.5em} | } 
\toprule
& \textbf{Baseline} & \textbf{0\%} & \textbf{0.001\%} & \textbf{0.005\%} & \textbf{0.01\%} & \textbf{0.05\%} & \textbf{0.1\%} & \textbf{0.5\%} & \textbf{1\%} & \textbf{5\%} & \textbf{10\%} & \textbf{20\%} & \textbf{50\%} & \textbf{100\%} \\
\midrule
  \hline
  Avg [kpps] & 927.77 & 924.54 & 926.15 & 926.53 & 919.92 & 905.18 & 893.01 & 798.66 & 750.51 & 563.83 & 437.74 & 289.55 & 106.68 & 85.79 \\ 
  \hline
  SD [kpps] & 2.129 & 2.764 & 2.955 & 2.804 & 4.874 & 1.977 & 1.401 & 4.067 & 2.857 & 3.959 & 1.712 & 0.640 & 0.296 & 0.203 \\ 
  \hline
  CV & 0.23\% & 0.30\% & 0.32\% & 0.30\% & 0.53\% & 0.22\% & 0.16\% & 0.51\% & 0.38\% & 0.70\% & 0.39\% & 0.22\% & 0.28\% & 0.24\% \\ 
  \hline
  CI$_{95}$ / Avg & 0.14\% & 0.18\% & 0.19\% & 0.18\% & 0.32\% & 0.13\% & 0.09\% & 0.31\% & 0.23\% & 0.42\% & 0.23\% & 0.13\% & 0.17\% & 0.14\% \\ 
  \hline
\end{tabular}
\caption{Scapy Opt Collector.}
\end{table*}
\end{center}

\begin{center}
\begin{table*}
\begin{tabular}{ | m{4.5em} | m{1cm}| m{2.7em} | m{2.9em} | m{2.9em} | m{2.7em} | m{2.7em} | m{2.7em} | m{2.7em} | m{2.7em} | m{2.7em} | m{2.7em} | m{2.7em} | m{2.7em} | m{2.5em} | } 
\toprule
& \textbf{Baseline} & \textbf{0\%} & \textbf{0.001\%} & \textbf{0.005\%} & \textbf{0.01\%} & \textbf{0.05\%} & \textbf{0.1\%} & \textbf{0.5\%} & \textbf{1\%} & \textbf{5\%} & \textbf{10\%} & \textbf{20\%} & \textbf{50\%} & \textbf{100\%} \\
\midrule
  \hline
  Avg [kpps] & 1030.93 & 1014.06 & 1015.55 & 1009.69 & 1017.13 & 1017.46 & 1017.05 & 1011.04 & 1012.02 & 1027.22 & 1076.97 & 1163.53 & 1525.47 & 3353.94 \\ 
  \hline
  SD [kpps] & 3.979 & 2.331 & 2.115 & 4.348 & 2.215 & 1.450 & 3.711 & 3.599 & 2.551 & 1.828 & 3.571 & 2.140 & 2.194 & 27.545 \\ 
  \hline
  CV & 0.39\% & 0.23\% & 0.21\% & 0.43\% & 0.22\% & 0.14\% & 0.36\% & 0.36\% & 0.25\% & 0.18\% & 0.33\% & 0.18\% & 0.14\% & 0.82\% \\ 
  \hline
  CI$_{95}$ / Avg & 0.23\% & 0.14\% & 0.12\% & 0.26\% & 0.13\% & 0.09\% & 0.22\% & 0.21\% & 0.15\% & 0.11\% & 0.20\% & 0.11\% & 0.09\% & 0.49\% \\ 
  \hline
\end{tabular}
\caption{eBPF Collector (1 map).}
\end{table*}
\end{center}

\begin{center}
\begin{table*}
\begin{tabular}{ | m{4.5em} | m{1cm}| m{2.7em} | m{2.9em} | m{2.9em} | m{2.7em} | m{2.7em} | m{2.7em} | m{2.7em} | m{2.7em} | m{2.7em} | m{2.7em} | m{2.7em} | m{2.7em} | m{2.5em} | } 
\toprule
& \textbf{Baseline} & \textbf{0\%} & \textbf{0.001\%} & \textbf{0.005\%} & \textbf{0.01\%} & \textbf{0.05\%} & \textbf{0.1\%} & \textbf{0.5\%} & \textbf{1\%} & \textbf{5\%} & \textbf{10\%} & \textbf{20\%} & \textbf{50\%} & \textbf{100\%} \\
\midrule
  \hline
  Avg [kpps] & 1030.93 & 1017.78 & 1011.81 & 1012.46 & 1017.66 & 1011.95 & 1013.00 & 1013.51 & 1005.81 & 1024.31 & 1066.74 & 1140.63 & 1431.96 & 2636.05 \\ 
  \hline
  SD [kpps] & 3.979 & 3.446 & 2.992 & 1.913 & 1.424 & 2.577 & 1.035 & 3.242 & 3.300 & 1.462 & 1.438 & 3.343 & 4.074 & 2.318 \\ 
  \hline
  CV & 0.39\% & 0.34\% & 0.30\% & 0.19\% & 0.14\% & 0.25\% & 0.10\% & 0.32\% & 0.33\% & 0.14\% & 0.13\% & 0.29\% & 0.28\% & 0.09\% \\ 
  \hline
  CI$_{95}$ / Avg & 0.23\% & 0.20\% & 0.18\% & 0.11\% & 0.08\% & 0.15\% & 0.06\% & 0.19\% & 0.20\% & 0.09\% & 0.08\% & 0.18\% & 0.17\% & 0.05\% \\ 
  \hline
\end{tabular}
\caption{eBPF Collector (4 maps).}
\end{table*}
\end{center}